\documentclass{aa}  

\usepackage{natbib,twoopt}
\usepackage{multirow}
\usepackage{amsmath}
\usepackage{amssymb} 
\usepackage[T1]{fontenc}
\usepackage{blindtext}
\usepackage[colorlinks = true,
            linkcolor = blue,
            urlcolor  = blue,
            citecolor = blue,
            anchorcolor = blue]{hyperref}

\usepackage{graphicx}
\usepackage{gensymb}
\usepackage{subcaption}
\usepackage{pythonhighlight}
\usepackage{listings}
\usepackage{xcolor}
\usepackage{multirow}
\usepackage{siunitx}
\usepackage{natbib}
\usepackage{hyperref}

\usepackage{orcidlink}

\usepackage{amsmath} 

\usepackage[varg]{txfonts}
\usepackage{natbib} 

\usepackage{graphicx}	
\usepackage{amsmath}	

\definecolor{codegreen}{rgb}{0,0.6,0}
\definecolor{codegray}{rgb}{0.5,0.5,0.5}
\definecolor{codepurple}{rgb}{0.58,0,0.82}
\definecolor{backcolour}{rgb}{0.95,0.95,0.92}

\defcitealias{Tiengo2023}{T23}

\lstdefinestyle{mystyle}{
  backgroundcolor=\color{backcolour}, commentstyle=\color{codegreen},
  keywordstyle=\color{magenta},
  stringstyle=\color{codepurple},
  basicstyle=\ttfamily\footnotesize,
  breakatwhitespace=false,         
  breaklines=true,                 
  captionpos=b,                    
  keepspaces=true,                                               
  showspaces=false,                
  showstringspaces=false,
  showtabs=false,                  
  tabsize=2
}

\begin{document} 

\title{Probing the interstellar medium toward GRB 221009A through X-ray dust scattering}

\author{B. Vaia\inst{1,2,3}\orcidlink{0000-0003-0852-0257},
        Ž. Bošnjak\inst{5}\orcidlink{0000-0001-6536-0320},
        A. Bracco\inst{6,7}\orcidlink{0000-0003-0932-3140},
        S. Campana\inst{8}\orcidlink{0000-0001-6278-1576},
        P. Esposito\inst{1,3}\orcidlink{0000-0003-4849-5092},
        V. Jelić\inst{4}\orcidlink{0000-0002-6034-8610}
        A. Sacchi\inst{9}\orcidlink{0000-0002-7295-5661} and
        A. Tiengo\inst{1,3}\orcidlink{0000-0002-6038-1090}}
\authorrunning{B. Vaia et al.}

\institute{Scuola Universitaria Superiore IUSS Pavia, Piazza della Vittoria 15, 27100 Pavia, Italy\\ \email{beatrice.vaia@iusspavaia.it}
  \and Department of Physics, University of Trento, Via Sommarive 14, 38123 Povo (TN), Italy 
  \and INAF - Istituto di Astrofisica Spaziale e Fisica Cosmica di Milano, Via A. Corti 12, 20133 Milano, Italy
  \and Ruđer Bošković Institute, Bijenička cesta 54, 10000 Zagreb, Croatia
  \and University of Zagreb Faculty of Electrical Engineering and Computing, Unska ul. 3, 10000 Zagreb, Croatia
  \and  INAF – Osservatorio Astrofisico di Arcetri, Largo E. Fermi 5, 50125 Firenze, Italy
  \and Laboratoire de Physique de l’Ecole Normale Supérieure, ENS, Université PSL, CNRS, Sorbonne Université, Université de Paris, F-75005 Paris, France
 \and INAF - Osservatorio Astronomico di Brera, Via Bianchi 46, 23807 Merate (LC), Italy
 \and Center for Astrophysics | Harvard \& Smithsonian, 60 Garden Street, Cambridge, MA 02138, USA
 }
   \date{Received; accepted}

\makeatletter
\renewcommand*\aa@pageof{, page \thepage{} of \pageref*{LastPage}}
\makeatother
 
\abstract{
The observation of 21 X-ray dust-scattering rings around the extraordinarily bright gamma-ray burst (GRB) 221009A provides a unique opportunity to study the interstellar medium (ISM) through which the X-ray radiation traveled in our Galaxy and, by difference, in the host galaxy as well. 
In particular, since the ring intensity and radius at a given time depend on the amount of dust and on its distance, respectively {\it XMM-Newton} and {\it Swift} 
images allowed us to map the ISM around the direction of the GRB with better resolution than in the existing optical and infrared-based 3D dust maps, both in the plane of the sky (few arcminutes) and along the line of sight (from $\simeq 1$ pc for dust clouds within 1 kpc to $\simeq 100$ pc for structures at distances larger than 10 kpc).
As a consequence, we can revise prior estimates of the GRB soft X-ray fluence, obtaining a $\sim$35\% lower value, which, however, still indicates a substantial excess with respect to the extrapolation of the spectral models constrained by hard X-ray observations.

Additionally, we detect significant spectral variability in two azimuthal sectors of the X-ray rings, which can be fully attributed to different Galactic absorption in these two directions. The comparison of the total hydrogen column density inferred from spectral fitting, with the Galactic contribution derived from the intensity of the X-ray rings, in the same sectors, allowed us to more robustly constrain the absorption in the host galaxy to $N_{\rm{H,z=0.151}}= (3.7\pm0.3)\,\times\,10^{21}\,\rm{cm^{-2}}$. This result is relevant not only to characterize the ISM of the host galaxy and to understand how the GRB radiation might have affected it, but also to model the broad-band spectrum of the GRB afterglow and to constrain the properties of a possible underlying supernova.
}


\maketitle
\nolinenumbers
\section{Introduction}\label{intro}

GRB 221009A is the brightest gamma-ray burst ever recorded \citep{Burns2023} and occurred in a direction close to the Galactic plane (b = 4\degree.32, l = 52\degree.96), resulting in the formation of 21 X-ray rings due to scattering by interstellar dust in our Galaxy (\citealt{Tiengo2023}, hereafter \citetalias{Tiengo2023};
\citealt{Vasilopoulos2023, Williams2023,zhao24}).

In the Milky Way, dust comprises only about 1\% of the interstellar medium (ISM) total mass \citep{Bohlin1978}, while the majority is gas, found in both atomic and molecular forms. Despite its small mass fraction, dust has a significant impact on astrophysical processes in all galaxies. Dust grains contribute to cooling effects that influence star and planet formation (see, e.g., \citealt{Schneider2002, johansen2017}) and provide surfaces for chemical reactions, supporting the formation of complex molecules. Dust has also a strong impact on many astronomical observations. First, it causes the extinction of radiation, mainly in the optical and UV wavebands, which, for extragalactic objects, requires a correction accounting for dust both in our Galaxy and in the host galaxy.
Second, the thermal emission of interstellar dust contaminates astronomical observations 
at long wavelengths, significantly affecting, in particular, 
the study of the Cosmic Microwave Background (CMB). 

In X-ray astronomy, dust impacts observations by absorbing and scattering X-ray photons. Although X-ray absorption is dominated by the contribution of gas metals in the ISM, high resolution spectroscopy makes it possible to isolate the contribution of some dust constituents or to infer their presence from the abundances of specific elements, such as Iron or Oxygen (see, e.g., \citealt{psaradaki23}).
When only low-resolution X-ray spectra are available, the quantity of dust along the line of sight can be estimated by modeling the photoelectric X-ray absorption and making assumptions on the dust-to-gas ratio.

Dust scattering of X-rays from bright point sources is a well-known and thoroughly studied phenomenon. For steady sources, this scattering produces a diffuse halo around the source at soft X-ray energies \citep{Premitt,Smith2008}. However, if the source undergoes a distinct flare followed by a period of quiescence, the scattering instead appears as discrete rings \citep{Vaughan2004,Tiengo2010, Heinz2016, Pintore2017} which expand over time.
The angular size of the formed expanding ring, $\theta$, is given by the relation:
\begin{equation}
\label{eq_theta1}
\theta = \sqrt{\frac{2c \Delta t}{D_s} \left(1 - \frac{D_d}{D_s}\right)},
\end{equation}
where $c$  is the speed of light, $\Delta t$ is the time delay between the direct and scattered X-rays, $D_s$ is the distance to the X-ray source, and $D_d$ is the distance to the dust cloud \citep{Miralda}.
In the case of GRBs scattered by Galactic dust, the distance to the X-ray source, $D_s$, is much greater than the distance to the dust layer, $D_d$. Under these circumstances, the term $D_d/D_s$ becomes very small, and the expression for the angle $\theta$ simplifies to:
\begin{equation}
\label{eq_theta2}
\theta [\rm{arcmin}]= 4.455\sqrt{\frac{\Delta t[\rm{days}]}{D_s[\rm{kpc}]}}.
\end{equation}
The flux of the ring, as a function of energy ($E$), due to the single scattering of GRB photons in a thin dust cloud within our Galaxy, and observed to expand from $\theta_1$ to $\theta_2$, can be described as:
\begin{equation}
\label{spettro}
F_{\rm{ring}}(E) = \frac{F(E)}{T_{\rm{exp}}}\Delta N_{\rm{H}}\,\sigma_{\theta_{1,2}}(E),
\end{equation}
where $F(E)$ is the GRB specific fluence, $T_{\rm{exp}}$ is the observation duration, $\Delta N_{\rm{H}}$ is the equivalent hydrogen column density of dust in the cloud averaged over the entire ring, and $\sigma_{\theta_{1,2}}(E)$ is the scattering cross section integrated between $\theta_1$ and $\theta_2$.
Therefore, observing X-ray rings allows us to study both the properties of the source and the intervening dust.
\citetalias{Tiengo2023} derived the fluence and spectrum of the soft X-ray emission of GRB 221009A from the observations of its dust-scattering rings, performed few days after the GRB with the European Photon Imaging Camera (EPIC; \citealt{pncamera,MOScamera}) onboard the \textit{X-ray Multi Mirror (XMM)-Newton} satellite. Here we use the same dataset, supported by observations with the X-ray Telescope (XRT; \citealt{XRT}) onboard the \textit{Swift Neil Gehrels Observatory} \citep{Gehrels2004} to study the distribution of interstellar dust responsible for scattering, both along the line of sight and in the plane of the sky. 
GRB 221009A presents a unique opportunity to investigate the Galactic interstellar medium through X-ray scattering and absorption due to the combination of its exceptional fluence and its ideal position in Galactic coordinates. A  GRB of such luminosity occurring at such a close distance is estimated to be a once-in-10,000-years event \citep{Burns2023} and its line of sight intercepts all the Milky Way spiral arms up to a distance of nearly 20 kpc from us (\citealt{Vasilopoulos2023, Williams2023}; \citetalias{Tiengo2023}). 
The paper is organized as follows: Section 2 details the reduction of the observations, Section 3 outlines the analysis, Section 4 discusses the results, and Section 5 presents our conclusions.

\section{Observations and Data reduction}\label{sec:1}
In this study, we examined \textit{Swift}/XRT and \textit{XMM-Newton} observations of GRB~221009A, performed between October 10 and October 14. The associated observation IDs are presented in Table 1. 
In later time observations, only the rings produced by the most distant clouds were fully covered (see Eq.~\ref{eq_theta2}), 
while here we mostly focus on clouds within 800 pc, which can also be identified through optical extinction.
On the other side, \textit{Swift} exposures taken within 1 day from the GRB discovery were not considered because the windowed-timing mode was used and no 2D images were produced.

\begin{table*}[ht]
	\centering
	\caption{Log of the \textit{\textit{Swift}/XRT} and \textit{XMM-Newton} observations of GRB 221009A.}
	\label{tab:1}
	\begin{tabular}{lcccc} 
		\hline
		Satellite/Instrument & Obs. ID & Start time (UTC) & Stop time (UTC) & Net Exp. time(ks)\\
		\hline
        \textit{Swift}/XRT &  01126853004 & 2022-10-10T08:05:13 & 2022-10-10T17:23:57 & $3.0$\\
        \textit{Swift}/XRT &  01126853005 & 2022-10-10T18:31:37 & 2022-10-10T19:02:39 & $1.5$\\
        \textit{Swift}/XRT &  01126853006 & 2022-10-10T20:22:49 & 2022-10-11T13:50:34 & $13.0$\\
        \textit{XMM-Newton}/MOS2 &  0913991501 & 2022-10-11T20:52:49 & 2022-10-12T10:46:09 & $50.0$\\
        \textit{Swift}/XRT &  01126853008 & 2022-10-12T01:15:43 & 2022-10-12T21:54:53 & $3.0$\\
        \textit{Swift}/XRT &  01126853009 & 2022-10-13T00:42:50 & 2022-10-09T04:22:52 & $4.4$\\
        \textit{XMM-Newton}/MOS2 & 0913991601 & 2022-10-14T05:41:21 & 2022-10-14T22:52:21 &  $34.0$\\
    	\hline
	\end{tabular}
\end{table*}
\subsection{XMM-Newton}\label{sec:1.1}
\textit{XMM-Newton} observed GRB 221009A for 50 ks on 2022 October 11 (Obs1) and for 62 ks on 2022 October 14 (Obs2). For minimizing pileup in Obs1 the EPIC pn \citep{pncamera} and the two EPIC Metal Oxyde Semiconductor (MOS; \citealt{MOScamera}) cameras were operated in Timing mode, with the thick optical-blocking filter. In Timing mode, only the peripheral CCDs of the MOS cameras offer full imaging capabilities from $\sim 5^\prime$ to $\sim 15^\prime$, from the target position. In Obs2 all the EPIC cameras were in full-frame mode and the thin optical filter was used.
The second part of this observation was affected by strong and variable particle background. Therefore, we limited the analysis to the longest uninterrupted time interval with quiescent background, resulting in a net exposure time of 33.5 ks for the MOS2 and 29.7 ks for the pn. For both observations, we do not report the analysis of the MOS1 data due to the permanent damage sustained by two external CCDs during the early stages of the mission.\\
\textit{XMM-Newton} observed the GRB three additional times on October 30, November 1, and November 11 2022 (Observation IDs: 0913991701, 0913991801, and 0913991901). In these observations, only the rings produced by the most distant clouds ($D_d > 3$~kpc) are visible, and therefore 
we used them only 
for subtracting the background in the analysis reported in Appendix~\ref{appA}.\\
To subtract the contribution from the X-ray background as well as from the GRB afterglow and its possible dust-scattering halo in Obs2, without any contamination from X-ray rings, we searched for archival observations of extragalactic point sources with similar flux and  spectrum, and behind a significant amount of Galactic dust.  As our best choice, we selected an observation of the bright high-redshift quasar RBS 315 (Observation ID: 0690900201) performed in January 2013, at a similar phase of the Solar Cycle, in order to have a similar contamination from the unfocused particle background \citep{GastaldelloSP}. This quasar is located in a relatively dusty region ($A_{\rm{v}} = 0.7$; \citealt{Schlafly2011}) and was observed by \textit{XMM-Newton} in full-frame mode with the thin filter, as in Obs2 (see Appendix~\ref{appB} for further details).\\
The data were processed using Science Analysis Software (SAS) 20.0.0 \citep{2004ASAS} and the latest calibration files. The EPIC events were cleaned with standard filtering expressions\footnote{For pn data: \texttt{$\#$XMMEA\_EP$\&\&$(FLAG==0)$\&\&$(PATTERN<=4)} and for MOS: \texttt{$\#$XMMEA\_EM$\&\&$(PATTERN<=12).}}. To maximize the signal-to-noise ratio of the rings, the analysis was confined to the 0.7–4 keV energy band, as in \citetalias{Tiengo2023}.
Point-like sources were removed by excluding circular regions after running the SAS source detection task \texttt{emldetect}.
For the spectral analysis response matrices were generated using the SAS tasks \texttt{rmfgen} and \texttt{arfgen}.

\subsection{Swift-XRT}\label{sec:1.2}
In this work, we analyzed \textit{Swift} observations of GRB 221009A made with XRT from October 10 to October 13, all taken in photon-counting mode. The corresponding observation IDs are listed in Table 1. Additionally, we used late-time observations of the GRB performed in March and April 2023 (observation IDs from 01126853076 to 01126853084) to create a single sky background event file. We also employed observations from October 25–29, 2022 (Observation IDs 01126853023 to 01126853027), where the GRB afterglow was still present but the rings were sufficiently large and faint, to isolate the afterglow contribution (see Appendix~\ref{appB} for details).
The data were processed with standard procedures using the FTOOLS task {\tt xrtpipeline}. 
Point-like sources were removed by excluding circular regions after source detection (we chose to exclude point-like sources with a signal to noise ratio > 3).
For the spectral analysis the Redistribution Matrix File (RMF) was obtained from the calibration database released on 01-01-2013 and the Ancillary Response Files (ARFs) were created using the task \texttt{xrtmkarf} (imposing \texttt{psfcorr=no}).
\section{Data analysis and results}\label{sec:2}

\begin{figure}
	\includegraphics[width=\columnwidth]{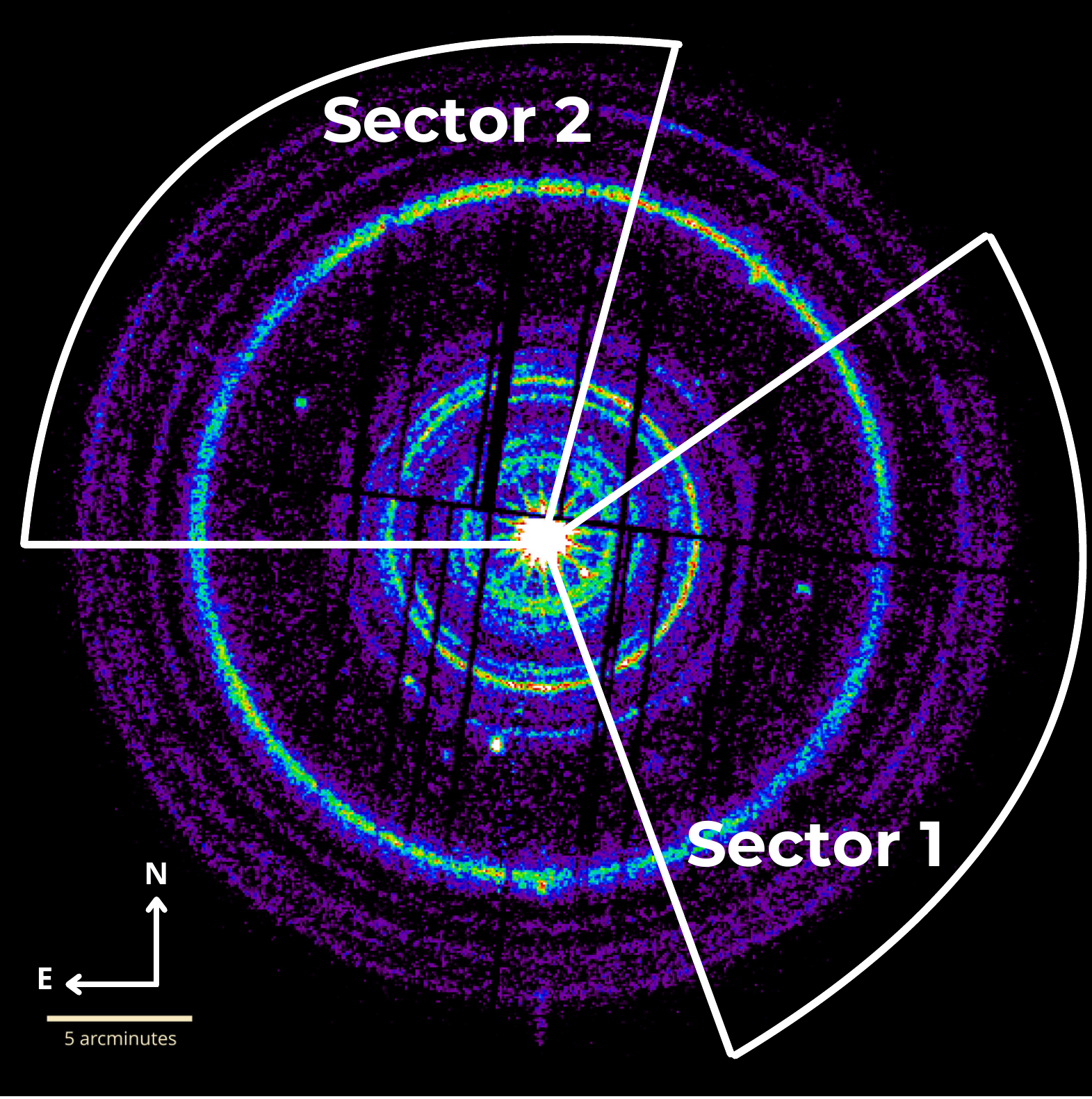}
    \caption{\textit{XMM-Newton} combined X-ray observations of GRB 221009A, with the two sectors used for our analysis highlighted.} 
    \label{fig_set}
\end{figure}

Since \textit{Swift}/XRT began observing GRB 221009A in photon-counting mode one day after the burst, it could observe rings formed by dust clouds at heliocentric distances smaller than 300 pc, which were already outside the EPIC field of view (FOV) during the \textit{XMM-Newton} observations (see Eq.~\ref{eq_theta2}). Unfortunately, due to the telescope's off-axis pointing, we were not able to observe these rings across the entire \textit{Swift}/XRT FOV, but only within a 105° wide sector ($290\degree-395\degree$, measured counterclockwise from the West direction) which extended up to $14^\prime$ from the GRB.
This sector, that we will call Sector 1 (see Fig. \ref{fig_set}), represents therefore the only region in which we can map the dust distribution throughout the whole Galaxy (from $\sim 0.1$ to $\sim 20$ kpc).
As noted in \citetalias{Tiengo2023}, there are significant variations in the azimuthal distribution of the surface brightness of the brightest X-ray rings,
which can be explained by the non-uniform distribution of dust in the intervening clouds. To exploit this azimuthal variation, we selected another sector with the same size  as Sector 1 (Sector 2; see Fig \ref{fig_set}).
The selection of this sector was driven by the analysis of X-ray absorption in 
Obs2 (see Appendix~\ref{appA}), which showed that Sector 2 ($75\degree-180\degree$) is the only region with a significant $N_{\rm{H}}$ excess. This result is confirmed also by the total $N_{\rm{H}}$ map of the GRB~221009A region estimated from \textit{Planck} submm data of thermal dust emission (see Fig.4 in \citetalias{Tiengo2023}).

\subsection{Galactic Dust Distribution}\label{sec:2.1}

In Sectors 1 and 2, we analyzed the expanding dust-scattering rings using the pseudo-distance distribution \citep{Tienghetti2006}, which we briefly summarize in the following. For each detected event, we computed the 
time delay of the photons respect to the GRB time ($t_{\rm{i}}$) and the angle between the scattered photon and the source direction ($\theta_{\rm{i}}$):

\begin{equation}
    t_{\rm{i}} = T_{\rm{i}} - T_0,
\end{equation}

\begin{equation}
     \theta_{\rm{i}}^2 = (x_{\rm{i}} - X_{\rm{GRB}})^2 + (y_{\rm{i}} - Y_{\rm{GRB}})^2,
\end{equation}
where $x_{\rm{i}}$, $y_{\rm{i}}$, and $T_{\rm{i}}$ are the detector coordinates and the time of arrival of the i-th event, and $T_0$, $X_{\rm{GRB}}$, and $Y_{\rm{GRB}}$ are the GRB time (59861.55568 MJD) and detector coordinates, respectively.
Using these new coordinates, we computed the pseudo-distance defined as:
\begin{equation}
\label{eq_distance}
D_{\rm{i}} = 2ct_{\rm{i}} / \theta_{\rm{i}}^2 = 827\,t_{\rm{i}}\, [s] \,\theta_{\rm{i}}^{-2}\,[\rm{arcsec}]\, \rm{pc} 
\end{equation}
For events which are not related to dust-scattered photons, this quantity does not represent a true physical distance. However, for photons from a ring which expands according to Eq. \ref{eq_theta2}, the pseudo-distance, $D_{\rm{i}}$, corresponds to the distance of the dust grains that scattered them. Consequently, all the X-rays scattered by a given dust cloud are detected at times and positions corresponding to approximately the same values of $D_{\rm{i}}$, resulting in a peak in the pseudo-distance distribution at the cloud's distance.

\begin{figure*}
\centering
\subfloat{\includegraphics[width = 0.45 \textwidth]{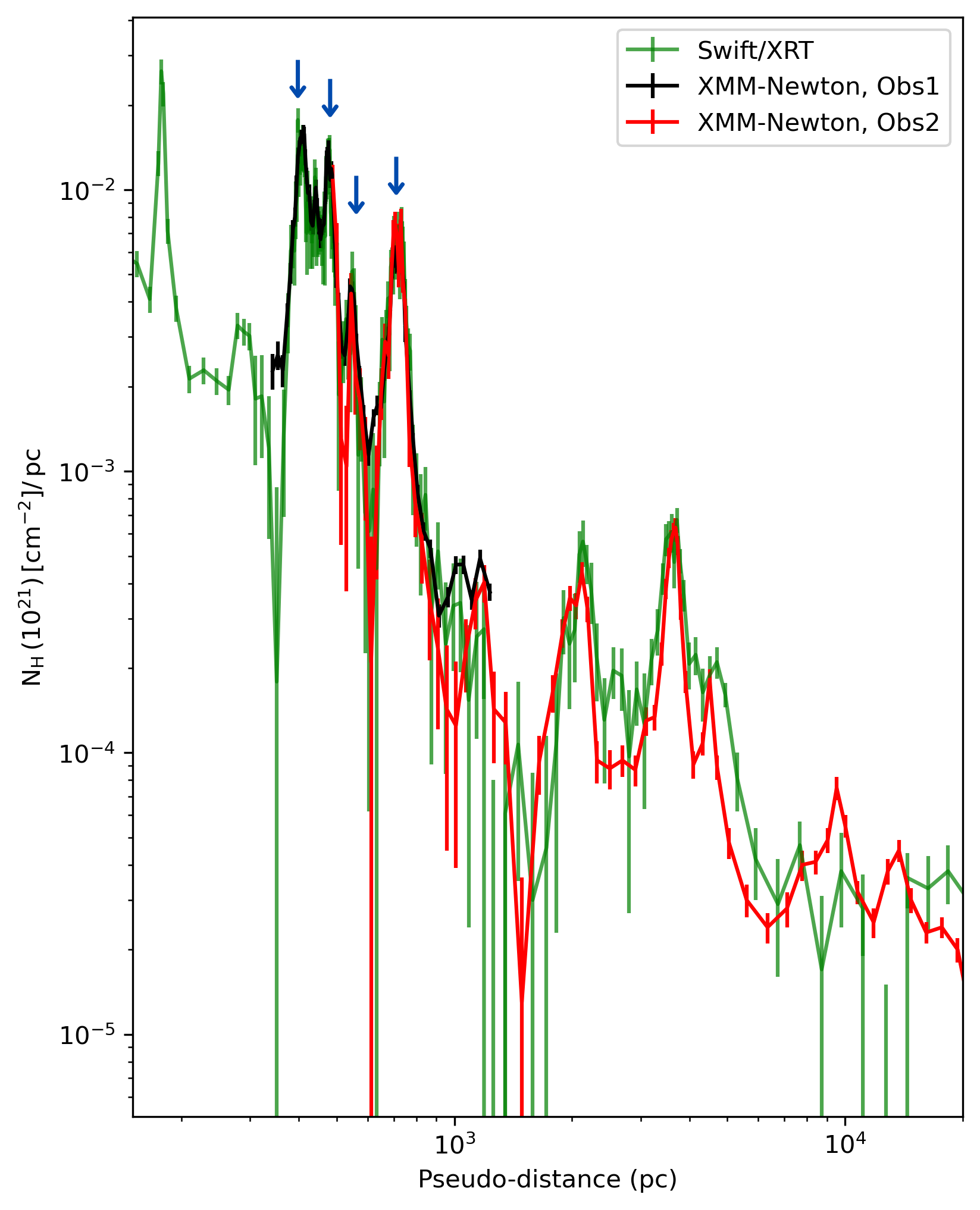}} \label{nh_dif_settore1}
\subfloat{\includegraphics[width = 0.45 \textwidth]{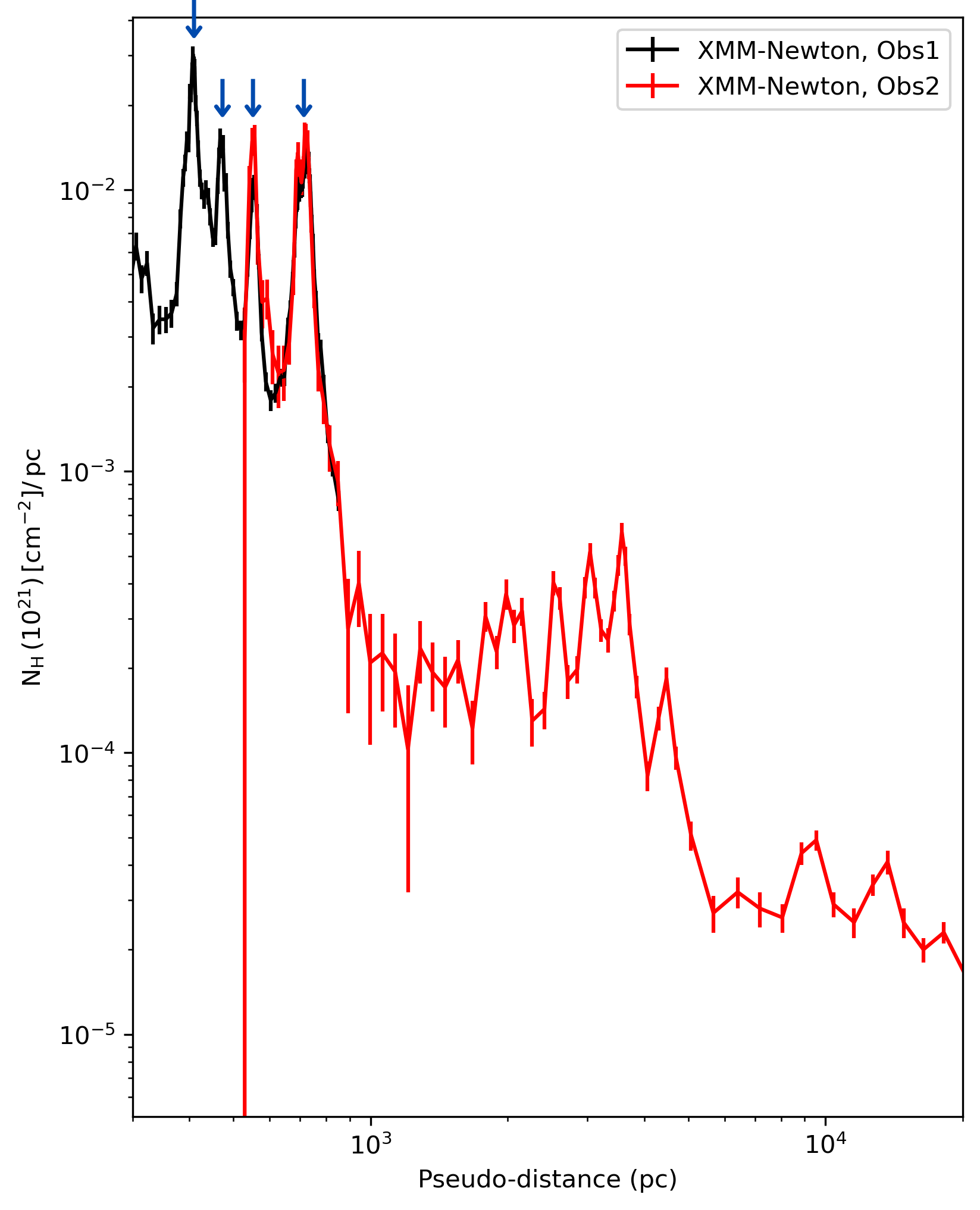}} \label{nh_dif_settore2}
\caption{Left panel: Differential Hydrogen column density in Sector 1; in green \textit{Swift}-XRT data, in black  the data obtained with the first \textit{XMM-Newton} observation and in red the data obtained with the second \textit{XMM-Newton} observation. Right panel: Differential Hydrogen column density in Sector 2; in black  the data obtained with the first \textit{XMM-Newton} observation and in red the data obtained with the second \textit{XMM-Newton} observation. The blue arrows highlight the four main peaks observed by both \textit{XMM-Newton} and \textit{Swift} satellites.}
   \label{nh_dif}
\end{figure*}

After constructing the histogram of the pseudo-distances for each sector (already presented in \citetalias{Tiengo2023} for the whole FoV) and subtracting the background (see Appendix~\ref{appB}), we transformed this count distribution $C(D)$ into the equivalent hydrogen column density of the dust $\Delta N_{\rm{H}}(D)$. This parameter can be obtained from our spectral model because the dust model is normalized to the number of hydrogen atoms \citep{2004Zubko}. For this transformation we used a conversion factor $f(\theta, t)$ that 
must be computed separately for every angular distance from the GRB position ($\theta$) and for each observation, accounting for the time elapsed since the GRB ($t$), the exposure duration and the detector efficiency. The procedure we used to compute such conversion factors is described in Appendix~\ref{appC}.
Applying these conversion factors, we obtained the differential  hydrogen column density distribution shown in Fig.~\ref{nh_dif} for the two sectors in the two MOS2 datasets and only for Sector 1 in the combined XRT observations (see Table~\ref{tab:1}), which allow us to sample nearby dust.
The $N_{\rm{H}}$ distributions obtained across the entire field of view for both MOS and pn cameras were compared and found to be consistent. However, in this work, we report the $\Delta N_{\rm{H}}$ maps only for the MOS2 camera, as it provides better coverage of Sector 1.
Fig.~\ref{nh_dif} demonstrates that the various datasets within the same sector are compatible with each other.

Due to the short duration of the GRB prompt emission, the brightness of the X-ray rings in different observations does not depend on the time variability of the source, but only on the spatial distribution of dust and on the angular dependence of the scattering cross-section, according to the selected dust model. To evaluate the azimuthal uniformity of 
the dust clouds at different angular distances ($\theta$), we generated the 
diagrams in Fig.~~\ref{fig:figura_inc}, which display the integrated $N_{\rm{H}}$ for the four 
main peaks in  Fig.~\ref{nh_dif}
observed by both the \textit{Swift} and \textit{XMM-Newton} satellites. In these charts, data for Sector 1 were represented as positive values of $\theta$, while the data for the other sector are shown as negative values. To reduce uncertainties, \textit{Swift} observations from multiple orbits were merged.
From these figures, it is evident that the clouds at 406 pc, 550 pc, and the cloud complex at 700 pc contain more dust in Sector 2 than in Sector 1. On the contrary, for the cloud at 475 pc, we do not detect any  significant variability.
The inset of the last panel in Fig. \ref{fig:figura_inc} illustrates the contributions of the two clouds within the cloud complex at $\sim$700 pc
in Sector 1: the cloud at 729 pc dominates up to $10^\prime$, but at larger angular distances, covered by the second \textit{XMM-Newton} observation, the corresponding X-ray rings fade away and the contribution from the cloud at 695 pc becomes predominant.

\begin{figure*}
\centering
\includegraphics[width=0.75\textwidth]{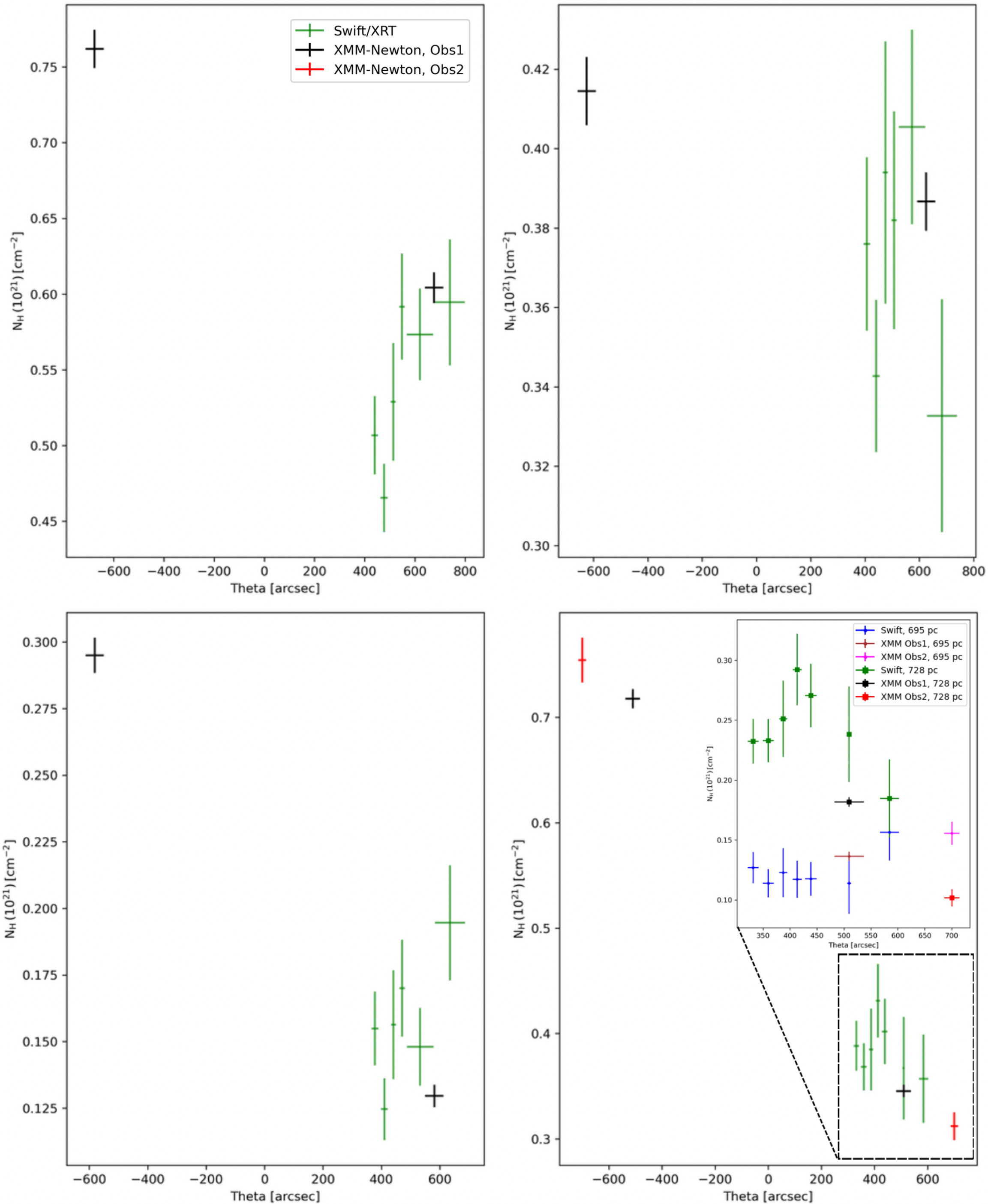}
\label{fig:figura_inc}
\caption{Hydrogen column density obtained by integrating the pseudo-distance distribution at the four main peaks in Fig. \ref{nh_dif} (highlighted with blue arrows), centered at 406.3 pc (upper left panel), 475.2 pc (upper right panel), 553.6 pc (bottom left panel) and $\sim$700 pc (bottom right panel). The inset of the last panel shows the two clouds around 700 pc as separate values: the circles refer to the peak at 695.4 pc (\textit{Swift}-XRT in blue, \textit{XMM-Newton} in brown for Obs1, and pink for Obs2) and the squares to the peak  centered at 728.6 pc (\textit{Swift}-XRT in green, \textit{XMM-Newton} in black for Obs1, and red for Obs2). In all the other panels the values from \textit{Swift}-XRT observations are shown in green, and from \textit{XMM-Newton} observations in black (Obs1) and red (Obs2).}\label{fig:figura_inc}
\end{figure*}

\subsection{Absorption in the host galaxy}\label{sec:2.2}
In addition to the variation in ring intensity across the two different sectors (discussed in Sect.~\ref{sec:2.1}), we also detect spectral differences, 
which, as already shown in Appendix~\ref{appA}, can be explained by a different amount of Galactic absorption. 
The combined analysis of the spatially-resolved intensity and absorption of the dust-scattered X-ray emission of GRB~221009A
enables us to distinguish between Galactic absorption and the contribution from the host galaxy. Specifically, the intensity of the X-ray rings allows for the quantification of dust along the line of sight within our Galaxy, while the X-ray absorption in the ring spectra is produced by the combined effect of the ISM both in our Galaxy and in the host galaxy.  By assuming a dust-to-gas ratio and subtracting these values, we can determine the amount of dust (and gas) present in the host galaxy.

To calculate the total X-ray absorption for each sector, we selected the annular regions with the largest signal-to-noise ratio. We extracted the ring spectra from the first \textit{XMM-Newton} observation within the angular ranges from 8$^{\prime}$ to 9$^{\prime}$ (covering the two rings produced by dust at distances of 695 and 728 pc) and from 9$^{\prime}$ to 12$^{\prime}$ (including three rings formed by dust at 406, 439, and 475 pc). For the second \textit{XMM-Newton} observation, we extracted the spectra in the angular range from  11$^{\prime}$ to 12$^{\prime}$ (containing the two rings generated by dust at 695 and 728 pc), utilizing data from both the MOS2 and pn cameras.
For each spectrum, the background was extracted from the same detector region in the RBS 315 observation.
We fit simultaneously the spectra of each ring with the XSPEC model:
$\textsc{constant}\,\times\,\textsc{TBabs}\,\times\,\textsc{ringscat}\,\times\,\textsc{pegpwrlw}$, 
where, at odds with the model described
in 
Appendix~\ref{appA}, \textsc{TBabs}\
includes the 
cumulative absorption from the ISM in our Galaxy and in the host galaxy.  The angles in the \textsc{ringscat} spectral component were fixed to the limits of each ring 
and all the other spectral parameters, except for 
$N_{\rm{H}}$ in \textsc{TBabs} 
and 
the value of the $\textsc{constant}$,\footnote{Since we are interested only in the spectral shape to constrain the absorption, we can ignore the brightness of the X-ray rings.} were linked to the same  values for the spectra of the two sectors.
With this model, we obtained a good fit: $\chi^2/\rm{dof} = 140.4/150$, corresponding to a null-hypothesis probability (nhp) of $96.5\%$ (for a more detailed description of the model and an example of the best-fit spectrum refer to \citetalias{Tiengo2023}).  
The best-fit photon index of the power-law component ($\Gamma = 1.38 \pm 0.05$) is perfectly consistent with the one derived, making similar assumptions, in \citetalias{Tiengo2023} for the full rings and, as anticipated by the preparatory work in Appendix~\ref{appA}, the absorption in the two sectors is significantly different: $N_{\rm{H,abs\,1}}\,=\,(9.6 \pm 0.2)\,\times\,10^{21}\,\rm{cm^{-2}}$ and $N_{\rm{H,abs\,2}}\,=\,(12.0 \pm 0.2)\,\times\,10^{21}\,\rm{cm^{-2}}$ for sector 1 and 2, respectively. This confirms that the spectral differences between the two sectors can be fully attributed to Galactic X-ray absorption and we do not find any evidence for a different scattering cross-section in the two sectors. Significant spectral differences would be expected in case of strong spatial variability of the dust population (see, e.g., the spectral parameters obtained in \citetalias{Tiengo2023} for different dust models) or for non-spherical grains \citep{Draine2006}.

\begin{figure*}[h]
    \centering
    \begin{subfigure}[b]{0.3\textwidth} 
        \centering
        \includegraphics[width=\textwidth]{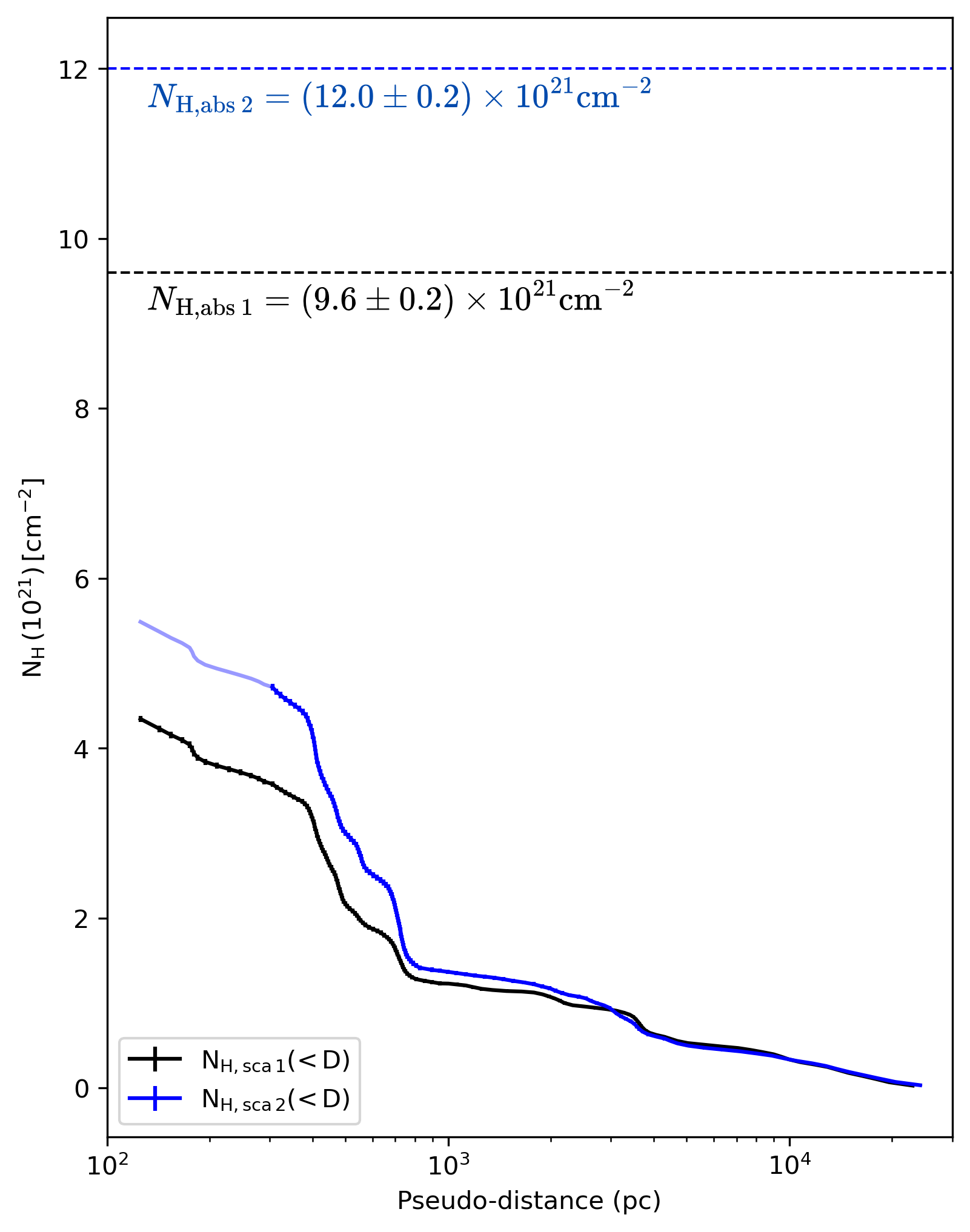}
        \label{cumulativo_nocorr}
    \end{subfigure}
    \hfill
    \begin{subfigure}[b]{0.3\textwidth} 
        \centering
        \includegraphics[width=\textwidth]{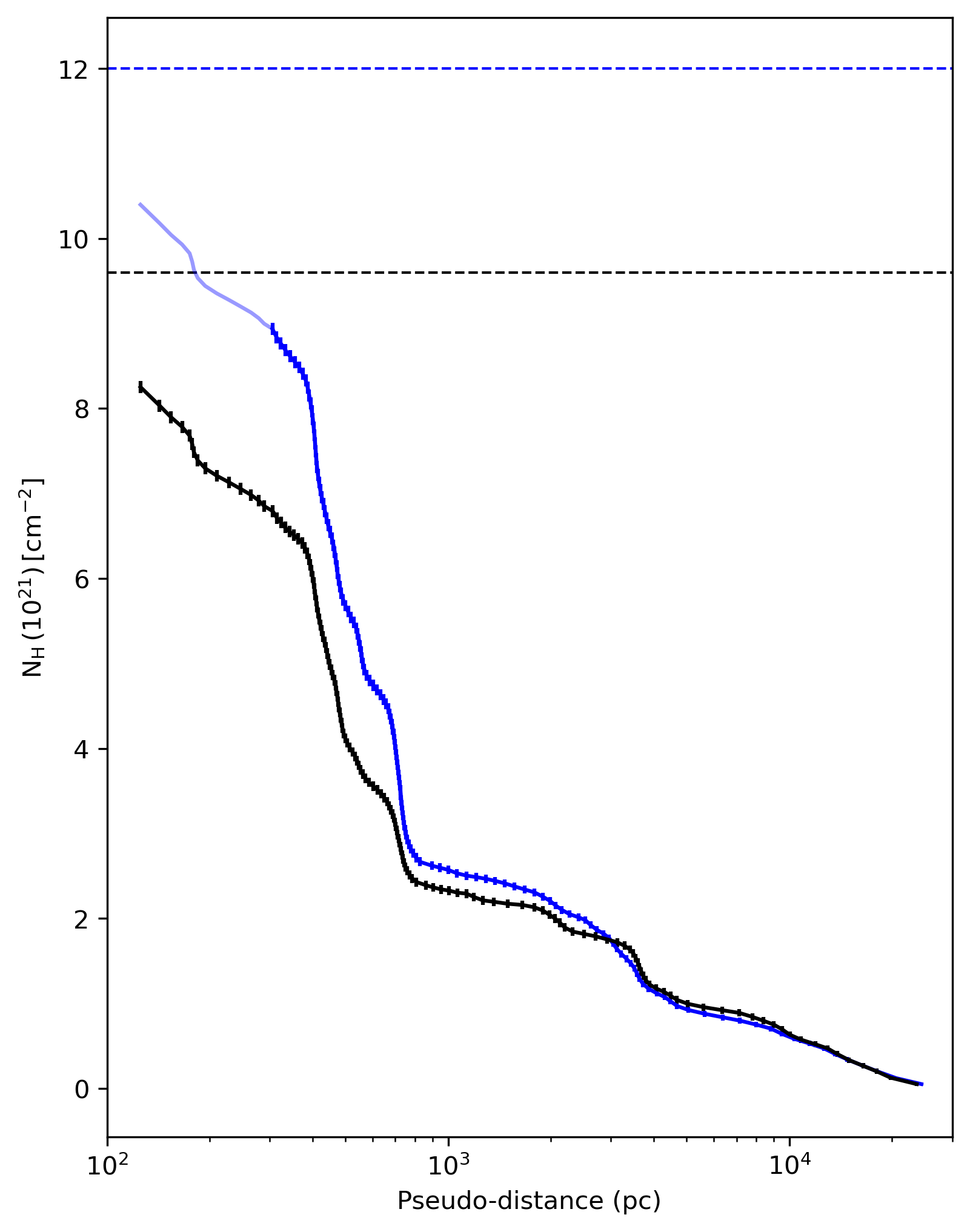}
        \label{cumulativo_corr1}
    \end{subfigure}
    \hfill
    \begin{subfigure}[b]{0.3\textwidth} 
        \centering
        \includegraphics[width=\textwidth]{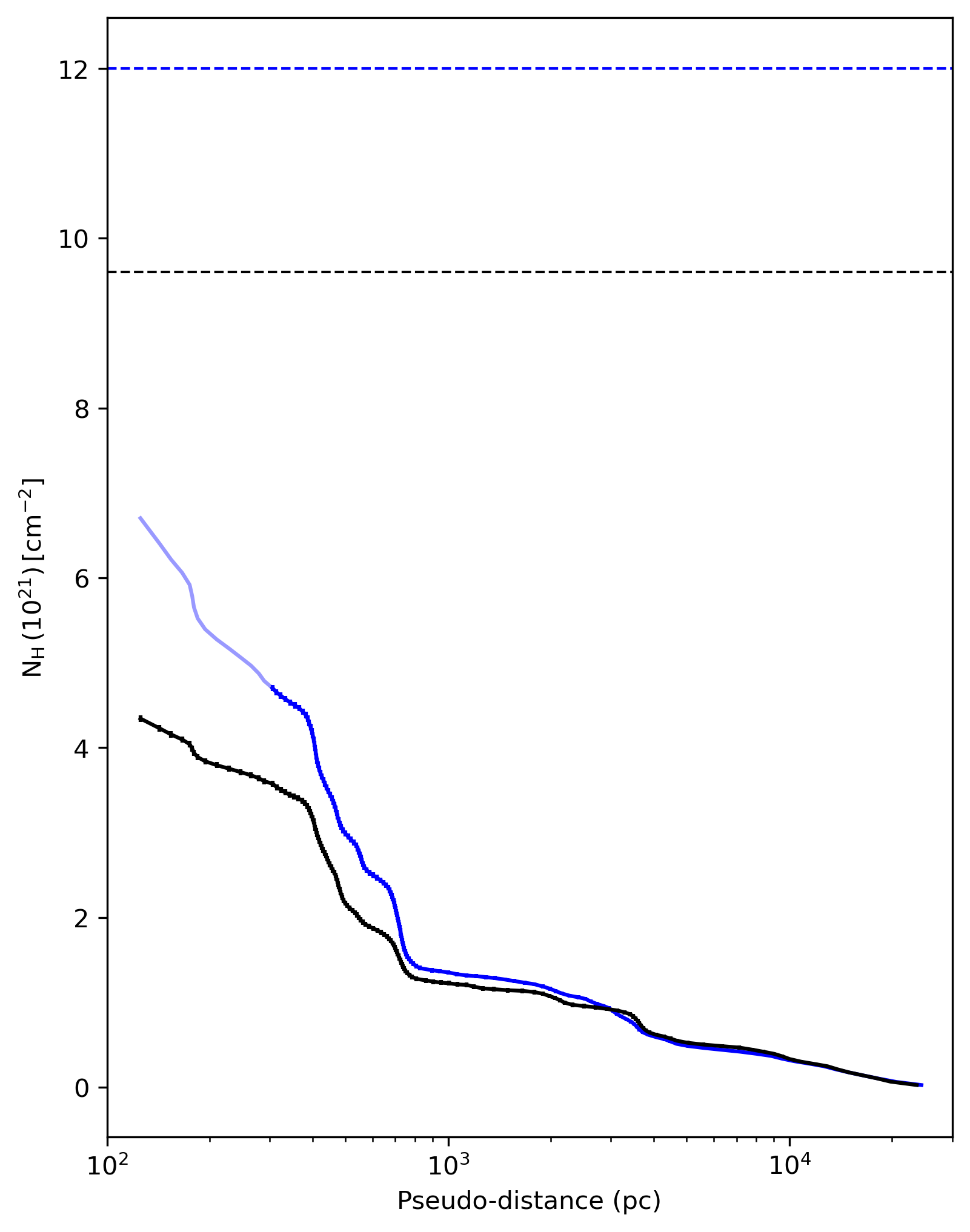}    
        \label{cumulativo_corr2}
    \end{subfigure}
\caption{Cumulative hydrogen column density in our Galaxy for Sector 1 (black solid lines) and Sector 2 (blue solid lines) derived from ring emission under different assumptions: GRB fluence from \citetalias{Tiengo2023} (left panel); 1.9 times lower fluence than in \citetalias{Tiengo2023} (middle panel); GRB fluence from \citetalias{Tiengo2023} and 2.6 times more dust in Sector 2 than in Sector 1 within 300 pc (right panel). In the first two panels, the same dust concentration is assumed for both sectors at distances < 300 pc. The dashed lines (black for Sector 1, blue for Sector 2) indicate the total hydrogen column density, including contributions from both our Galaxy and the host galaxy, measured from the absorption inferred from  X-ray spectral fitting.} 
\label{fig:cumulativo_tot}
\end{figure*}

For both sectors, we produced the cumulative $N_{\rm{H}}$ in our Galaxy from ring emission (left panel of Fig.~\ref{fig:cumulativo_tot}) by integrating the corresponding pseudo-distance distributions, starting from the largest distances. To complete the cumulative distribution for Sector 2, we assumed the same dust concentration as in Sector 1 for distances smaller than 300 pc. We utilized \textit{Swift} data up to 300 pc, Obs1 data up to 900 pc, and, beyond 900 pc, we used data derived from Obs2. The total integrated values for sector 1 and 2 are $N_{\rm{H,sca\,1}}=(4.34 \pm 0.04)\,\times\,10^{21}\,\rm{cm^{-2}}$ and $N_{\rm{H,sca\,2}}=(5.54 \pm 0.04)\,\times\,10^{21}\,\rm{cm^{-2}}$, respectively.

Although the difference between the final value of the cumulative dust distribution ($\Delta N_{\rm{H,sca}}=N_{\rm{H,sca\,2}}-N_{\rm{H,sca\,1}}$) and the total X-ray absorption ($\Delta N_{\rm{H,abs}}=N_{\rm{H,abs\,2}}-N_{\rm{H,abs\,1}}$) in the two sectors should be consistent, since it does not depend on the absorption in the host galaxy,
we found significantly different values: $\Delta N_{\rm{H,sca}}=(1.20 \pm 0.06)\,\times\,10^{21}\,\rm{cm^{-2}}$ and $\Delta N_{\rm{H,abs}}=(2.4 \pm 0.3)\,\times\,10^{21}\,\rm{cm^{-2}}$.

This discrepancy is likely due to some of the assumptions made during the computation of the cumulative dust map ($N_{\rm{H, sca}}$).  For instance, as described in Appendix \ref{appC}, we adopted a GRB fluence based on the values reported in \citetalias{Tiengo2023}. However, this measurement is affected by the systematic uncertainties discussed in Section \ref{sec:3.4}. Since  $N_{\rm{H, sca}}$ is inversely proportional to the GRB fluence (see Eq. \ref{spettro}), to reconcile the observed values, we could assume a 1.9 times lower GRB fluence. This adjustment results in the cumulative distributions shown in the central panel of Fig.~\ref{fig:cumulativo_tot} and an absorption in the host galaxy of $N_{\rm{H,z=0.151}}(1.8\pm0.3)\,\times\,10^{21}\,\rm{cm^{-2}}$.
The second assumption involves the extrapolation to small distances for the cumulative distribution of Sector 2, which is assumed to be the same as that of Sector 1. By instead assuming an excess of dust by a factor of 2.6 in the dust clouds at distances smaller than 300 pc in Sector 2, we obtain the cumulative dust distributions shown in the right panel of Fig. \ref{fig:cumulativo_tot}, and the resulting absorption in the host galaxy is $N_{\rm{H,z=0.151}}=(7.0\pm0.3)\,\times\,10^{21}\,\rm{cm^{-2}}$.
A more robust estimate of $N_{\rm{H,z=0.151}}$ can be derived from the detailed comparison of our dust distance distributions with 3D optical extinction maps, which will be discussed in the next section.

\section{Discussion}\label{sec:3}
We can compare the dust map derived from X-ray scattering with other maps generated in the direction of GRB~221009A using data at various wavelengths.  
Here we focus on the 3D extinction maps from \citet{Lallement2022}, \citet{Edenhofer2024}, and \citet{Green2019}.
\subsection{Comparison with the Lallement map}\label{sec:3.1}
 The extinction map constructed by \citet{Lallement2022} integrates \textit{Gaia} Early Data Release 3 (EDR3) and 2MASS photometric data with \textit{Gaia} EDR3 parallaxes. This map has a distance resolution of 25 pc and an angular resolution greater than \ang{1} in the plane of the sky\footnote{The map is publicly available on the EXPLORE website: \url{https://explore-platform.eu}.}.
 
\begin{figure*}
	\includegraphics[width=\columnwidth]{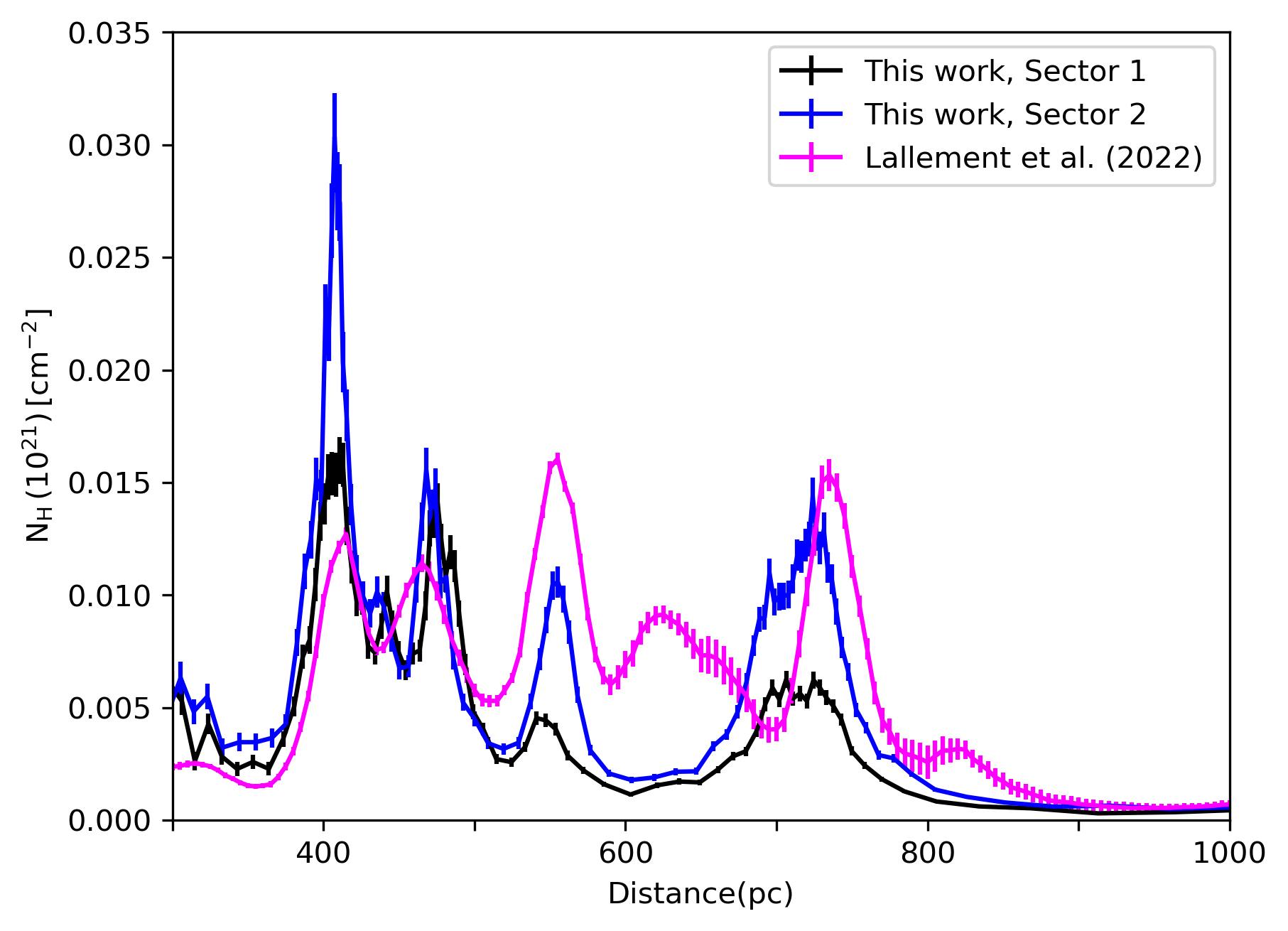}
    \includegraphics[width=\columnwidth]{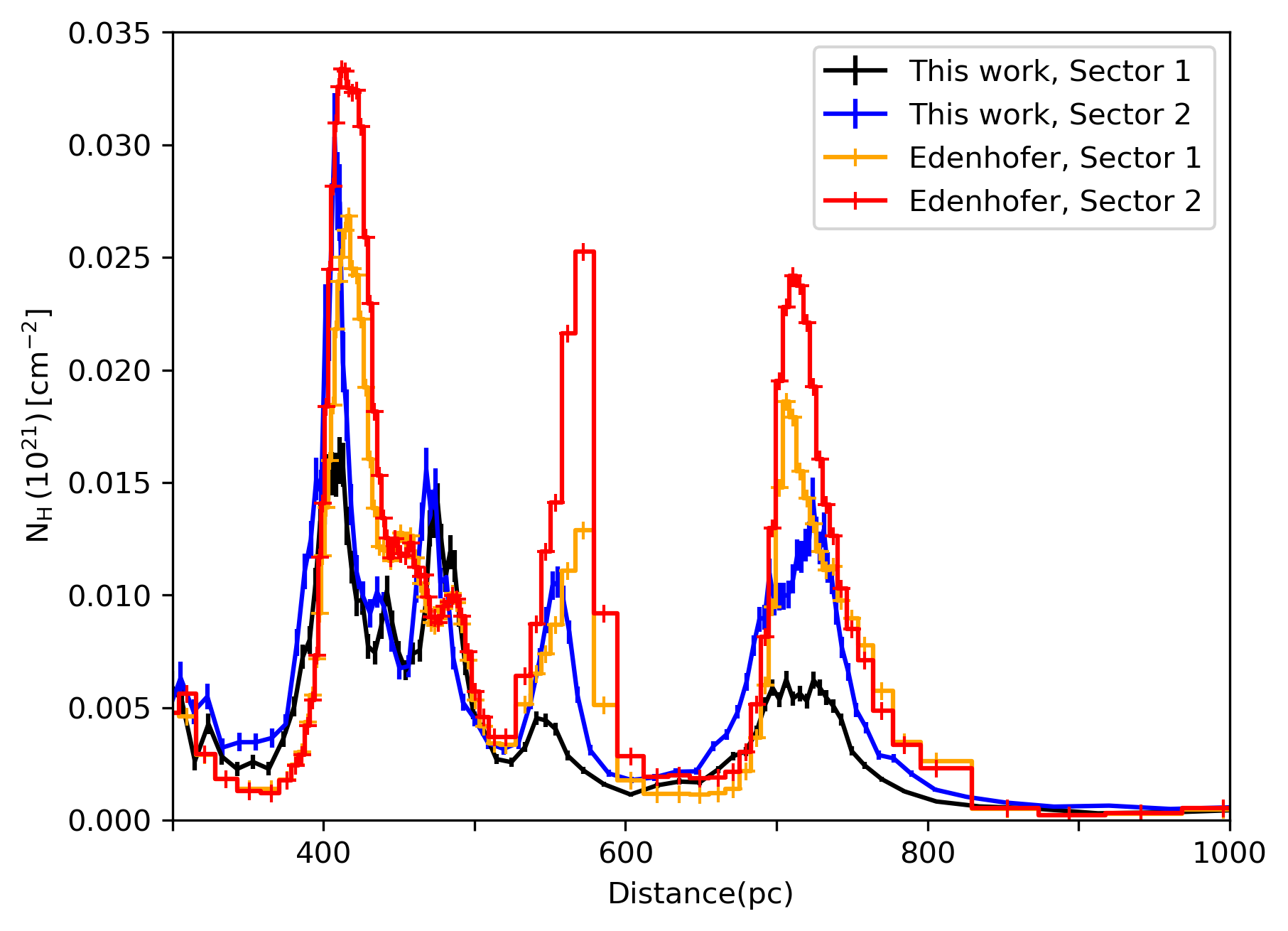}
    \caption{
    Differential hydrogen column density derived from X-ray dust scattering for Sector 1 (black) and Sector 2 (blue), compared to 3D optical extinction maps. In the left panel, the values derived by \citet{Lallement2022} are displayed in pink, while in the right panel the differential hydrogen column density from \citet{Edenhofer2024} is presented in orange for Sector 1 and in red for Sector 2. A conversion factor of $1.9\,\times\,10^{21}\rm{cm^{-2}\,\rm{mag^{-1}}}$ between $N_{\rm{H}}$ and $A_{\rm{v}}$ was used.} 
    \label{Lallement}
\end{figure*}

 Such angular resolution is not sufficient to separately probe the directions of the two sectors sampled by the X-ray data. However, most extinction increases in the \citet{Lallement2022} map extracted in the direction of GRB~221009A occur at the same distances as the peaks
of the pseudo-distance distributions of the two sectors (left panel of Fig. \ref{Lallement}).
This confirms the results reported by \citet{BarbaraPaper} for the dust distribution 
averaged over the full FOV.

Thanks to the conversion factors described in Appendix~\ref{appA} and adopting the $A_{\rm{v}}$ to $N_{\rm{H}}$ relation derived in \citetalias{Tiengo2023} ($N_{\rm{H}}/A_{\rm{v}} = 1.9\,\times\,10^{21}\rm{cm^{-2}\,\rm{mag^{-1}}}$), we can compare the peak heights as well.
Taking into account the lower angular resolution of the \citet{Lallement2022} map, the peak values are in good agreement. In fact, the extinction features that display the largest discrepancies correspond to peaks in the pseudo-distance distributions with the most prominent differences in the two sectors, which indicate a significant gradient in the dust spatial distribution of the associated cloud.
Moreover, features present in the \citet{Lallement2022} 
map but 
not detected through X-ray dust-scattering, such as the one at 650 pc, are likely due to dust clouds located at angular distances from the GRB position exceeding the radius of the largest X-ray ring ($\sim$12$^{\prime}$, corresponding to 2.3 pc at this distance) or to spurious structures generated by the interpolation procedure required by
the scarcity of suitable stars in the sampled volume \citep{Lallement2022}. 

\subsection{Comparison with the Edenhofer map}\label{sec:3.2}

The 3D dust differential extinction map presented by \citet{Edenhofer2024} is based on distance and extinction estimates for 54 million nearby stars, derived from \textit{Gaia} BP/RP spectra. This map offers improved angular resolution compared to the \citet{Lallement2022} map (up to $14^\prime$), with distance resolution varying from 0.4 pc at 69 pc to 7 pc at 1.25 kpc. It is publicly accessible through the \textsc{dustmaps} Python package \citep{Green2018}. The map provides unitless extinction values as defined in \citet{Zhang2023}. To convert it to Johnson’s V-band, we apply a factor of 2.8 \citep{Zhang2023}, and then convert it to $N_{\rm{H}}$ using the $A_{\rm{v}}/N_{\rm{H}}$ conversion factor from \citetalias{Tiengo2023}.

\begin{figure*}
	\includegraphics[scale=1.]{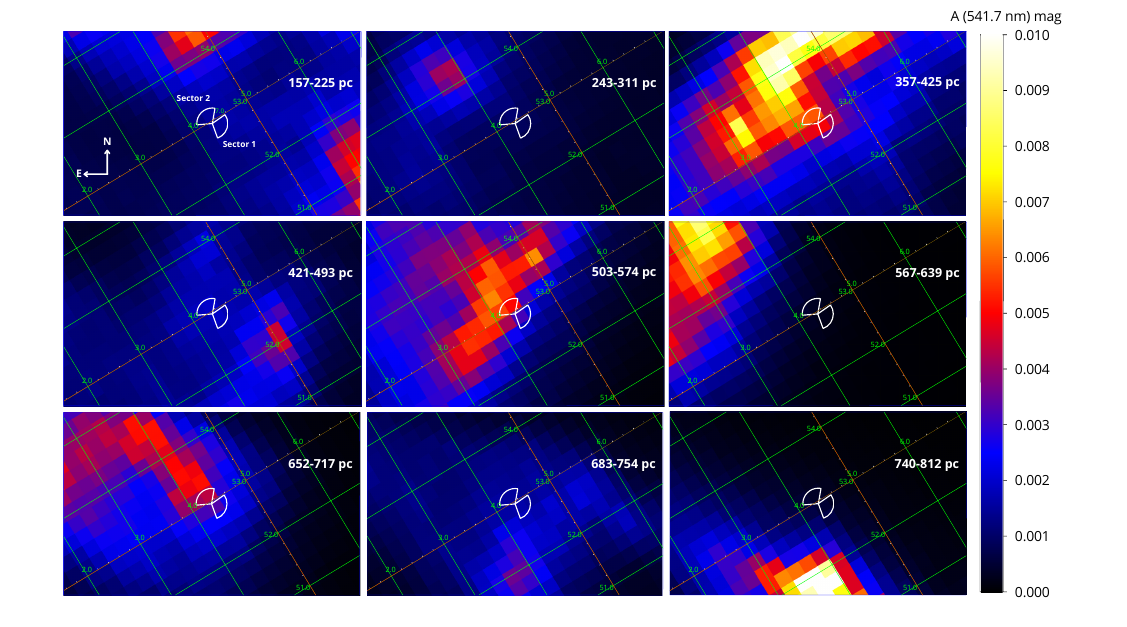}
    \caption{Extinction maps at $\lambda=541.7$~nm 
from \citet{Edenhofer2024} in the direction of GRB 221009A,
integrated in
the following distance intervals: 157-225 pc (top left panel), 243-311 pc (top central panel), 357-425 pc (top right panel), 421-493 pc (central left panel), 503-574 pc (central panel), 357-425 pc (central right panel), 652-717 pc (bottom left panel), 683-754 pc (bottom central panel), and 740-812 pc (bottom right panel). The regions covered by the two sectors studied in this work are highlighted in white in each panel. 
The green grid indicates the Galactic coordinates in degrees.}
    \label{Edenhofer_mappe}
\end{figure*}

Thanks to the better resolution of the \citet{Edenhofer2024} map, in this case we can extract the differential extinction map for Sector 1 ($\rm{RA = 288.10\degree}$, $\rm{Dec = 19.71\degree}$) and Sector 2 ($\rm{RA = 288.38\degree}$, $\rm{Dec = 19.90\degree}$) and compare them
with the corresponding pseudo-distance distributions (right panel of Fig. \ref{Lallement}). 
The three main peaks emerging in the \citet{Edenhofer2024} map are not only at compatible distances, but they also have similar relative heights, with systematically higher values in sector 2. The corresponding peak values, instead, are always larger in the \citet{Edenhofer2024} map, which suggests a lower fluence for the GRB prompt emission. The pseudo-distance distribution peaks at $\sim$440 pc and $\sim$480 pc are not resolved in the \citet{Edenhofer2024} map, which however confirms the spatial uniformity of dust at these distances.
Moreover, the \citet{Edenhofer2024} map does not detect the features around $\sim$600 pc and $\sim$800 pc, which are present in the \citet{Lallement2022} map but not in our pseudo-distance distributions. The images shown in Fig.~\ref{Edenhofer_mappe} 
allow us to confirm one of the hypothesis formulated above to interpret this discrepancy:
the broad bump above 600 pc is very likely caused by the dust cloud visible in the North-East direction
in the 567–639 pc distance bin (central right panel of Fig. \ref{Edenhofer_mappe}), which vanishes at $\sim$1$\degree{}$ from the GRB, well beyond the region explored by X-ray observations
(highlighted in white). Similarly, in the 740–812 pc bin (bottom right panel of Fig. \ref{Edenhofer_mappe}), another dust cloud outside the FOV of the X-ray instruments is probably responsible for the small bump detected by \citet{Lallement2022} above 800 pc (left panel of Fig. \ref{Lallement}).
Fig. \ref{Edenhofer_mappe} also presents the extinction maps in the distance bins where dust clouds are observed though X-ray scattering. These figures clearly show the morphology of such clouds in the sky plane and confirm that for most clouds Sector 2 contains more dust compared to Sector 1.

\subsection{Comparison with the Green map}\label{sec:3.3}

The \citet{Green2019} map combines stellar photometry from \textit{Gaia} Data Release 2 (DR2), Pan-STARRS 1, and 2MASS  to provide extinction estimates with \textit{Gaia} DR2 parallaxes. The map provides cumulative extinction across sight lines in 120 logarithmically spaced distance bins ranging from 63 pc to 63 kpc. In the direction of GRB~221009A, extinction can be measured for stars at distances from $\sim$0.3 to $\sim$5 kpc,
extending well beyond the volume covered by the \citet{Lallement2022} and \citet{Edenhofer2024} maps. The angular resolution varies between 3.4$^{\prime}$ and 13.7$^{\prime}$, depending on the region of the sky. This map is also publicly available via the Python package \textsc{dustmaps}. It provides extinction values in arbitrary units, which we converted to $A_{\rm{v}}$ using the conversion formula outlined in \citet{Green2019} (see Table 1 and Eq.~30).

\begin{figure*}
	\includegraphics[width=\columnwidth]{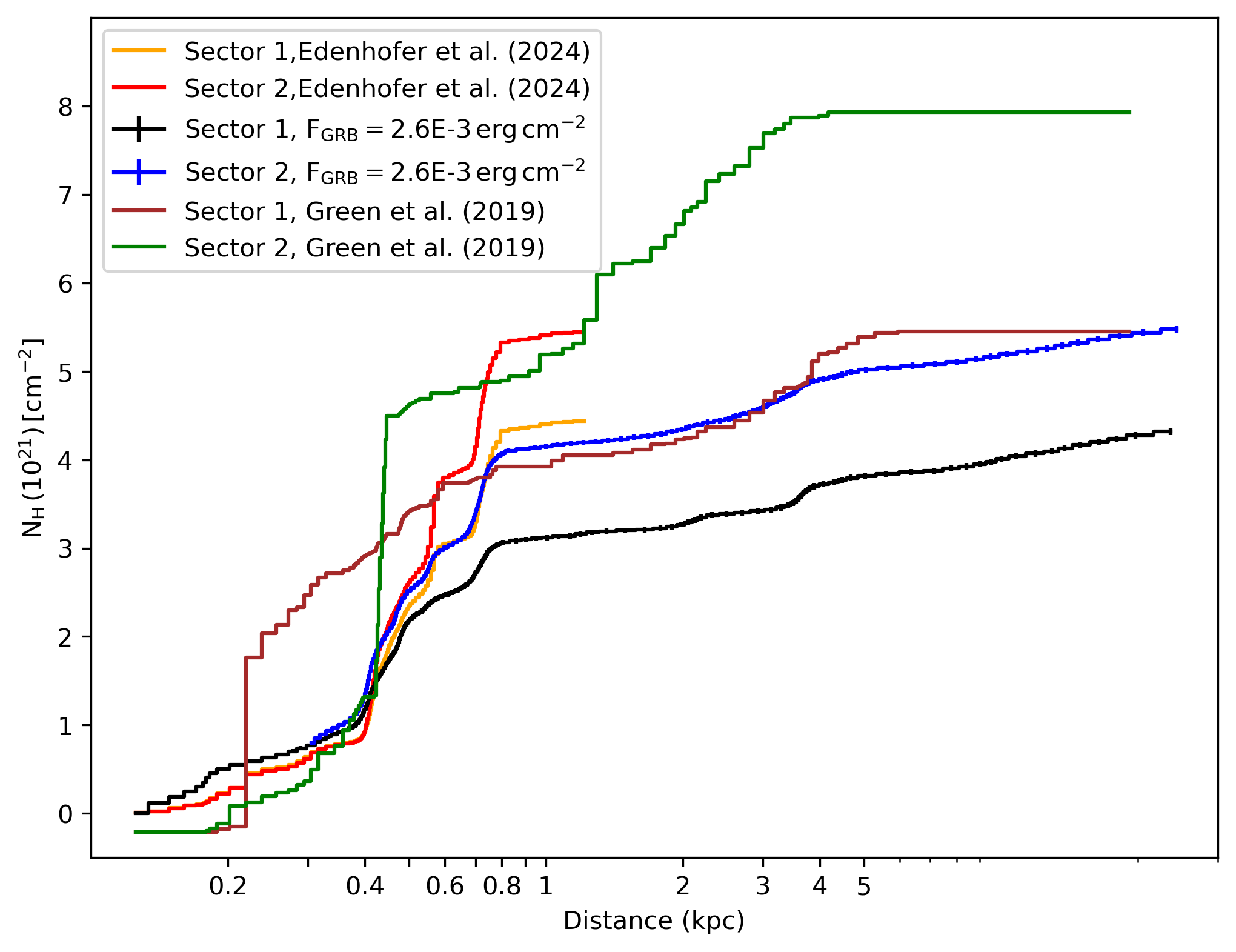}
 	\includegraphics[width=\columnwidth]{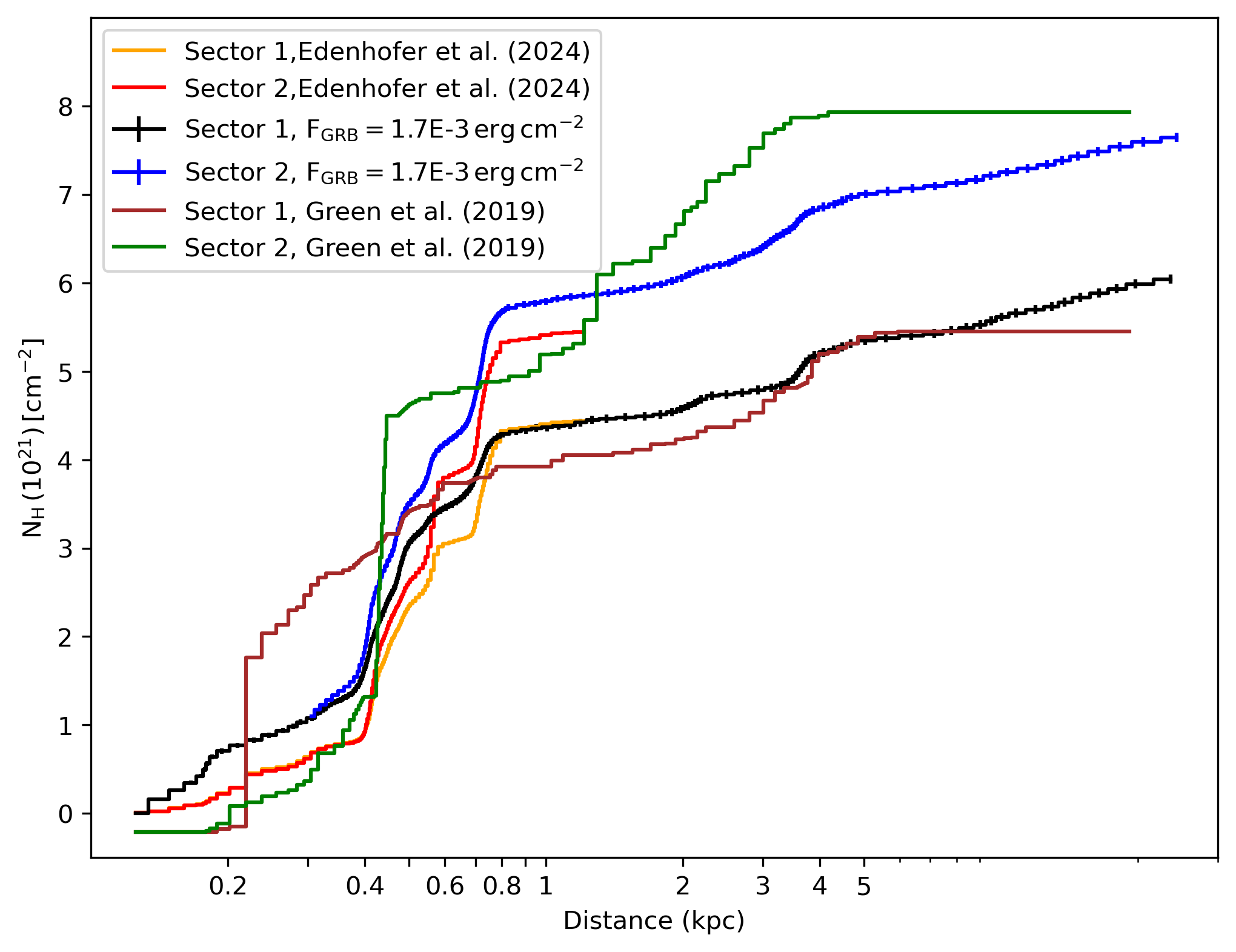}
    \caption{Cumulative hydrogen column densities obtained from X-ray dust scattering for Sector 1 (black) and for Sector 2 (blue), calculated using either the best-fit fluence from \citetalias{Tiengo2023} (left panel) or a factor 1.35 smaller fluence (right panel). In both cases the same dust distribution for distances <300 pc were assumed for the two sectors. Both panels also show a comparison with the cumulative hydrogen column density derived from \citet{Green2019} (brown for Sector 1  and green for Sector 2) and from \citet{Edenhofer2024} (orange for Sector 1 and red for Sector 2). A conversion factor between $N_{\rm{H}}$ and $A_{\rm{v}}$ equal to $1.9\,\times\,10^{21}\rm{cm^{-2}\,\rm{mag^{-1}}}$ was used.}
    \label{Green1}
\end{figure*}

We compared our pseudo-distance distributions in the two sectors with those from \citet{Green2019}, using the $A_{\rm{v}} - N_{\rm{H}}$ conversion factor from \citetalias{Tiengo2023}. The comparison was carried out by extracting the \citet{Green2019} maps of Sector 1 (brown in Fig. \ref{Green1}) and Sector 2 (green in Fig. \ref{Green1}) from the same directions chosen for the \citet{Edenhofer2024} maps in section 4.2.
As shown in the left panel of Fig. \ref{Green1}, the pseudo-distance distributions (Sector 1 in black and Sector 2 in blue) yield a total $N_{\rm{H}}$ along the line of sight within our Galaxy that is lower than the values found by \citet{Green2019} in the same sector.  As already discussed in Sect.~\ref{sec:3.2}, this discrepancy could result from our choice of the GRB fluence which, as explained in \citetalias{Tiengo2023}, is affected by large systematic uncertainties. For example, by decreasing the GRB fluence by a factor of 1.35, the cumulative $N_{\rm{H}}$ obtained from X-ray dust scattering in Sector 1 matches that of the Green map in the same sector at $\sim700$ pc and at 5 kpc and the one of \citet{Edenhofer2024} (reported in orange in Fig.\ref{Green1}) from 700 pc to 1 kpc. The total value of $N_{\rm{H,sca\,1}} = (6.07\pm 0.05)\,\times\,\,10^{21}\,\rm{cm^{-2}}$ is consistent with the average value provided by the \citet{Planck} 2D map in Sector 1: $N_{\rm{H}} = (6.3\pm 0.4)\,\times\,\,10^{21}\,\rm{cm^{-2}}$. 
The absorption value in the host galaxy found with this revised fluence is $N_{\rm{H,z=0.151}}=(3.7\pm0.3)\,\times\,10^{21}\,\rm{cm^{-2}}$. 
The difference $\Delta N_{\rm{H}}=2.48\,\times\,10^{21}\,\rm{cm^{-2}}$ between the values derived by \citet{Green2019} for the two sectors at 5 kpc (the farthest distance at which \citealt{Green2019} can estimate stellar extinction) is consistent with the  difference in X-ray absorption in the two sectors derived from the spectral analysis described in Sect.~\ref{sec:2.2} ($\Delta N_{\rm{H, abs}}=2.4 \pm 0.3\,\times\,10^{21}\,\rm{cm^{-2}}$). On the contrary, even assuming the reduced fluence, the excess of dust in Sector 2 derived from X-ray scattering is significantly smaller: $\Delta N_{\rm{H, sca}}= (1.63 \pm 0.07)\,\times\,10^{21}\,\rm{cm^{-2}}$. To achieve a compatible value, the dust content within 300 pc in Sector 2 should be $\sim$1.6 times greater than that in Sector 1 at the same distance.
Finally, we note that, as the X-ray absorption is primarily due to gas along the line of sight, these values could also be reconciled by assuming a different dust-to-gas ratio (see, e.g., \citealt{Zhu2017}).

\subsection{Impact on GRB 221009A properties}\label{sec:3.4}

Our last estimate of the X-ray absorption in the host galaxy, $N_{\rm{H,z=0.151}}=(3.7\pm0.3)\,\times\,10^{21}\,\rm{cm^{-2}}$, is significantly smaller than the value obtained by \citet{Williams2023} from the analysis of the afterglow X-ray spectrum, $N_{\rm{H,z=0.151}} = (1.4 \pm 0.4) \times 10^{22}\, \rm{cm^{-2}}$. This difference can be explained by the smaller value assumed by \citet{Williams2023} for the Galactic absorption ($N_{\rm{H,Gal}} = 5.38 \times 10^{21}\, \rm{cm^{-2}}$), derived from a $N_{\rm H}$ map with poor angular resolution \citep{Willingale2013}. As already noted in \citetalias{Tiengo2023}, the $N_{\rm{H}}$ inferred from \textit{Planck} data ($N_{\rm{H,Gal}} \simeq 9 \times 10^{21}\, \rm{cm^{-2}}$;  \citealt{Planck}) indicates a significantly greater absorption in our Galaxy at the afterglow position.

In the analysis of the multi-wavelength afterglow reported by \citet{kann23}, the impact of multiple values for the Galactic extinction, ranging from 0 to $A_{\rm{v}} = 5.2~\rm{mag}$ (corresponding to no additional extinction in the host galaxy), was explored.
\citet{Fulton2023} determined that in order to align the optical and X-ray fluxes observed 1-2 days after the burst, an additional extinction of $A_{\rm{v}} = 1.5~\rm{mag}$ ($N_{\rm{H}} = 2.85 \times 10^{21} \rm{cm^{-2}}$ using a conversion factor $N_{\rm{H}}/A_{\rm{v}} = 1.9\,\times\,10^{21}\rm{cm^{-2}\,\rm{mag^{-1}}}$; \citetalias{Tiengo2023}) is required. This excess is likely due to the contribution of the ISM in the host galaxy and it is only slightly smaller than our estimate, which corresponds to $N_{\rm{H}} = (3.5 \pm 0.3) \times 10^{21} \rm{cm^{-2}}$, if the correction due to the host galaxy redshift is not applied. 
Actually, \citet{Fulton2023} observed that this additional extinction decreased to
1.0 mag and 0.8 mag 2-3 days and 4-5 days after the GRB, respectively. The slightly larger X-ray absorption that we derive from the spectral analysis of the dust-scattered prompt GRB emission might therefore be part of this decreasing trend. A similar variability with time has been observed in other GRBs, such as GRB~190114C, and attributed to the photo-ionization of the surrounding ISM \citep{campana21}. 

The best estimate of the X-ray fluence of GRB 221009A in the soft X-ray range given in \citetalias{Tiengo2023} ($2.6 \times 10^{-3} \, \rm{erg \, cm^{-2}}$) is approximately an order of magnitude larger than the value obtained by extrapolating to the same energy range the GRB spectrum observed at higher energies \citep{Frederiks2023,An,Burns2023}. 
This estimate was derived assuming the amount of dust in the cloud at 0.73 kpc as determined by \citet{Lallement2022}, which, as discussed in Sect.~\ref{sec:3.1}, lacks adequate angular resolution. Decreasing the fluence by a factor 1.9 or 1.35,  as suggested by the spatially-resolved analysis reported in Sect.~\ref{sec:2.2} and~\ref{sec:3.3}, respectively, would only partially reduce the magnitude of the soft excess found in \citetalias{Tiengo2023}, which would still be a few times brighter than the extrapolation of the prompt GRB emission estimated from hard X-ray data.

The sector analysis reported here and the adoption of different extinction maps allow us to better control systematic effects, but our results are still based on a single dust model (BARE-GR-B) and a specific relation between hydrogen column density and optical extinction ($N_{\rm{H}} =1.9\,\times\,10^{21}\,A_{\rm{v}}\,\rm{cm^{-2}\,\rm{mag^{-1}}}$; \citetalias{Tiengo2023}). Even if the BARE-GR-B model is the best one describing the dust responsible for the X-ray rings observed around GRB 221009A, \citetalias{Tiengo2023} have shown that other dust models provide similarly good fits to the ring spectra (e.g., COMP-GR-B from \citealt{2004Zubko} or the \citealt{draine03} model). In these cases, we obtain a GRB fluence that can be different by up to a factor 1.5. 
Additionally, the $N_{\rm{H}}/A_{\rm{v}}$ relationship is also subject to systematic uncertainties, as demonstrated by the scatter in previous measurements across various sky regions and astrophysical objects, which range from  $1.0\,\times\,10^{21}$ to $2.8\,\times\,10^{21}\,\rm{cm^{-2}\,\rm{mag^{-1}}}$ \citep{Zhu2017}. Finally, 
although the spatially-resolved spectral analysis of the X-ray rings reported in Appendix~\ref{appA} suggests
that all the dust clouds are formed by dust grains with the same properties, 
we cannot exclude the existence of different dust populations depending on the local environment.

\section{Conclusion}

We presented a detailed analysis of the dust-scattered X-ray emission from GRB 221009A observed by \textit{XMM-Newton} and \textit{Swift}/XRT. The main aim of this analysis, supported by the comparison with recently published 3D extinction maps, was to constrain the properties of the ISM in the GRB direction both in our Galaxy and in the host galaxy. In particular, we generated a 3D map of the Galactic dust distribution in a $\sim$0.5$\degree$ wide region centered at the position of the GRB with a better resolution than previous studies, both in the plane of the sky (few arcminutes) and along the radial direction (approximately 1 pc for distances $D<1$ kpc and 100 pc for $D>10$ kpc). Although this map covers a very small sky region, the dust cloud distances are directly measured throughout the whole Galaxy using a geometrical method, in analogy with parallaxes for stars, and can therefore be used to calibrate other maps based on indirect or model-dependent distance measurements.

In addition, the spectral analysis of the brightest X-ray rings revealed considerable variability in the hydrogen column density of the total absorbing matter across the FOV, 
which, coupled to the correlated intensity variability of the dust-scattered radiation, allowed us to separate Galactic absorption from the host galaxy contribution. Such measurement is crucial not only to characterize the GRB local environment and its time evolution induced by the burst radiation, but also to constrain the intrinsic luminosity of a possibly associated supernova (see, e.g., \citealt{Shrestha2023}).

Finally, this work refined the soft X-ray fluence estimate from \citetalias{Tiengo2023} ($2.6 \times 10^{-3} \, \rm{erg \, cm^{-2}}$), reducing it by a factor of at least 1.35. However, the revised values, based on the comparison of our modeling of the dust-scattering rings with three different 3D extinction maps, are not small enough 
to eliminate the soft excess in the prompt X-ray emission found in the previous work.


\begin{acknowledgements}
This publication was produced while B.V. attending the PhD program in  in Space Science and Technology at the University of Trento, Cycle XXXVIII, with the support of a scholarship financed by the Ministerial Decree no. 351 of 9th April 2022, based on the NRRP - funded by the European Union - NextGenerationEU - Mission 4 "Education and Research", Component 1 "Enhancement of the offer of educational services: from nurseries to universities” - Investment 4.1 “Extension of the number of research doctorates and innovative doctorates for public administration and cultural heritage”. AB acknowledges the financial support of INAF through the initiative “IAF Astronomy Fellowships in Italy” (grant name MEGASKAT). SC acknowledge support from ASI under contract ASI/INAF I/004/11/6.

\end{acknowledgements}

\bibliography{main}{}

\begin{appendix}
\section{\\Azimuthal Variability of X-ray absorption}\label{appA}
\begin{figure}
	\includegraphics[width=\columnwidth]{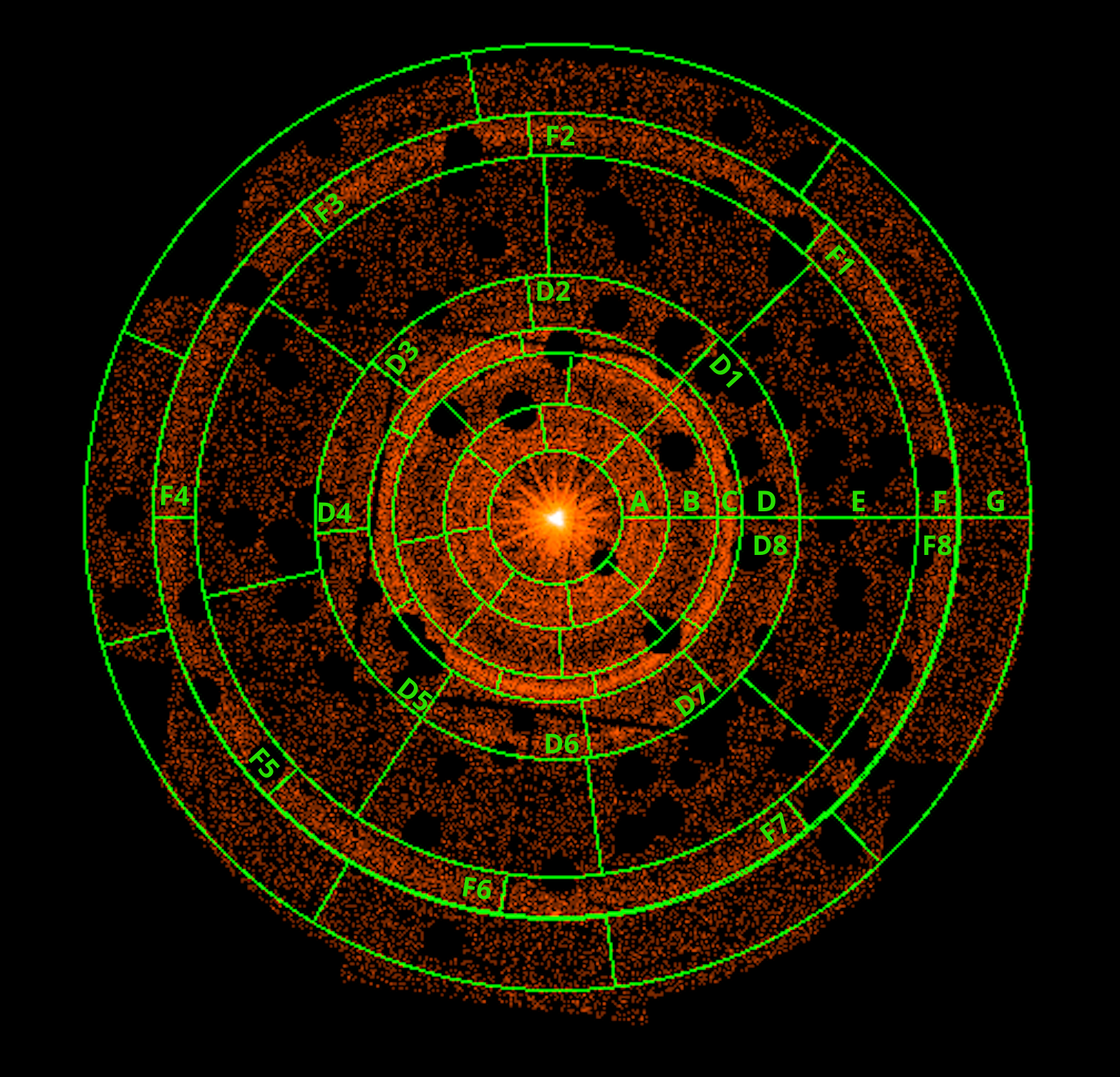}
    \caption{Annular sectors in which we performed the spectral analysis to evaluate the variation of X-ray absorption. The inner and outer radii for each ring are: Ring A ($2^\prime$–$3.37^\prime$), Ring B ($3.37^\prime$–$4.88^\prime$), Ring C ($4.88^\prime$–$5.62^\prime$), Ring D ($5.62^\prime$–$7.27^\prime$), Ring E ($7.27^\prime$–$10.85^\prime$), Ring F ($10.85^\prime$–$12.1^\prime$), and Ring G ($12.1^\prime$–$14.2^\prime$). Each ring is divided into 8 sectors containing the same number of counts. } 
    \label{preliminaryAnalysis1}
\end{figure}

\begin{figure}
	\includegraphics[width=\columnwidth]{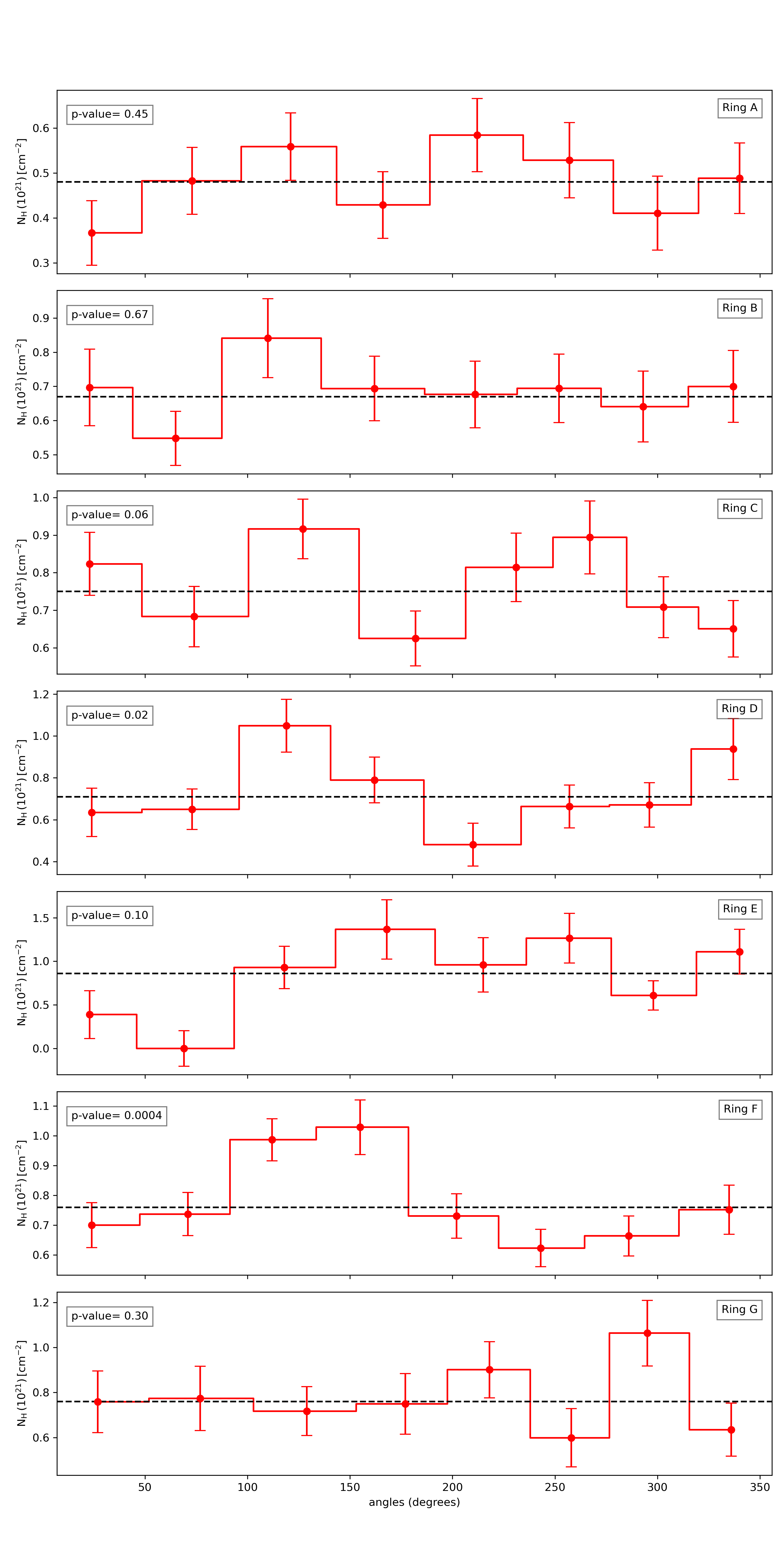}
    \caption{X-ray absorption for the seven rings derived from the spectral analysis using the model: $\textsc{constant}\,\times\,\textsc{zTBabs}\,\times\,\textsc{TBabs}\,\times\,\textsc{ringscat}\,\times\,\textsc{pegpwrlw}$.  Each sector within the same region is fit with a constant, yielding acceptable fits (p-value displayed in the upper left of each plot) for all rings except for ring F.
    } 
    \label{preliminaryAnalysis2}
\end{figure}

To optimize the selection of azimuthal sectors for mapping the Galactic dust in the directions around the GRB 221009A position, a spectral analysis of annular sectors with similar numbers of counts was performed to evaluate the possible variation of X-ray absorption.
The MOS2 field of view in Obs2 was divided into seven concentric annuli (labeled rings A through G) with adjacent scattering rings grouped together to form regions containing at least 8000 counts (Fig.\ref{preliminaryAnalysis1}). Each ring was then further subdivided into eight sectors, ensuring that all sectors within the same ring had about the same number of counts (>1000, to maintain sufficient statistics). The background spectra were obtained from the same region but extracted from the late time observation made with \textit{XMM-Newton} on November 11. 
The spectra of each annular sector were simultaneously fit using the XSPEC model already adopted in \citetalias{Tiengo2023}:
$\textsc{constant}\,\times\,\textsc{zTBabs}\,\times\,\textsc{TBabs}\,\times\,\textsc{ringscat}\,\times\,\textsc{pegpwrlw}$. 
In this model, \textsc{ringscat} takes into consideration the product $\Delta N_{\rm{H}}\,\sigma_{\theta_{1,2}}(E)$ in Eq.~\ref{spettro} and the prompt GRB spectrum is modeled as a power law (\textsc{pegpwrlw}) modified by absorption in both our Galaxy (\textsc{TBabs}) and the host galaxy (\textsc{zTBabs}). The model \textsc{ringscat} is based on the exact Mie calculation of the scattering cross section on spherical grains \citep{Mie}, with composition and size distribution according to the \textsc{BARE-GR-B} model \citep{2004Zubko}. In this dust model, that provides the best-fit to the spectra of the 19 X-ray rings analyzed in \citetalias{Tiengo2023}, the grains are bare (uncoated), composed of graphite, silicate and polycyclic aromatic hydrocarbons, and B star abundances are assumed \citep{2004Zubko}. 
All the parameters were fixed to the best-fit values found in \citetalias{Tiengo2023}, except for the Galactic absorption and the overall normalization (\textsc{constant}). 
An acceptable fit ($\chi^2/\rm{dof} = 2207/2177$; nhp = $32\%$) was obtained, which means that the angular dependence of the scattering cross-section (modeled by the \textsc{ringscat} spectral component) and a spatially-variable Galactic absorption can fully explain the spectral variability detected across the MOS2 FOV.
The best-fit Galactic absorption hydrogen column densities derived for each spectrum
are shown in Fig.~\ref{preliminaryAnalysis2}.
Ring F, which contains the X-ray radiation scattered by dust at $\sim$$700\,\rm{pc}$, displays
significant azimuthal variability, with a prominent absorption excess in sectors F3 and F4, which range from
$94\degree{}$ to $180\degree{}$. The level of X-ray absorption in the other sectors of this ring, $N_{\rm{H}}$=$0.726\,\pm\,0.018$, is compatible with the constant values derived for the other rings with radii larger than $\sim$3$^{\prime}$. For Ring A, instead, we found a lower column density, but, at such small angles, both the subtraction of the X-ray afterglow and the calculation of the scattering cross-section are affected by larger systematic uncertainties.


\section{Background subtraction in pseudo-distance distributions}\label{appB}

The subtraction of the contribution of events not produced by X-ray dust scattering to the pseudo-distance distribution is crucial to avoid overestimating the quantity of dust in our Galaxy. To analyze the \textit{Swift}/XRT observations, we implemented a double background-subtraction technique to isolate the emission from the rings by removing contributions from both the sky and instrumental background and the GRB afterglow.
First, we constructed a pseudo-distance distribution for each sector from late-time observations (Obs IDs from 01126853076 to 01126853084). These background distributions were generated by randomizing the time within the range of the \textit{Swift}/XRT observations of the GRB during which the rings were visible and by measuring the angular distance from the GRB in sky coordinates.

To further remove the afterglow, we used a second set of observations taken about 15 days after the burst (Obs IDs from 01126853023 to 01126853027). By this time, the afterglow was still bright, but the flux of the X-ray rings was below the detection threshold. The pseudo-distance distribution for the afterglow in each sector was produced as for the late time observations, which, in turn, were used to obtain a background-subtracted afterglow distribution. Such distributions were then enhanced by a factor corresponding to the ratio of the average afterglow flux in the 0.3 - 5 keV energy band measured during the observations from 10 to 13 October (when the rings were detected) and from 25 to 29 October (from which we extracted the pseudo-distance distribution of the afterglow). To obtain the net contribution from the X-ray dust scattering halo, we therefore subtracted from the original pseudo-distance distributions of the two sectors both these afterglow distributions and those obtained for the background at late times. 

In the two \textit{XMM-Newton} observations, we did not utilize late-time observations as background, due to the residual contamination by X-ray dust scattering. Instead, we subtracted both the sky background and the afterglow contribution using a single observation of a different target (RBS 315; see Section 2 for details).

For the first \textit{XMM-Newton} observation (Obs1), the subtraction was straightforward, involving a direct subtraction between the two pseudo-distance distributions, where angular distances were calculated in detector coordinates and the arrival times of the events in the background observation were randomly attributed in the time interval covered by Obs1. On the contrary, the subtraction in the second \textit{XMM-Newton} observation (Obs2) was complicated by the presence of an anomalous electronic noise in some MOS2 CCDs. Specifically, the ratio between the counts in individual CCDs and the total counts in the 5-10 keV band (where the X-ray rings are not detectable) showed an excess with respect to Filter Wheel Closed (FWC) exposures\footnote{The EPIC CCD cameras on board \textit{XMM-Newton} are equipped with a filter wheel system and 6 different filter setups, including a closed position, in which the sky is blocked by a $\sim\,1$ mm thick layer of Aluminum. The FWC exposures, which are dominated by the instrumental background, are typically used to model and subtract the internal instrumental background of the \textit{XMM-Newton} EPIC CCD cameras.}. For Obs1 and the background observation, instead, the count ratios of every CCD were consistent with those observed  in the FWC exposures. 

To account for this anomalous background in Obs2, we rescaled the background pseudo-distance distribution of each CCD before performing the subtraction. The rescaling factor was the ratio between the 5-10 keV counts in Obs2 and those in the background observation for each CCD. 

\section{Calculation of the conversion factors for pseudo-distance distributions}\label{appC}

To compute the conversion factors $f(\theta, t)$, which allow us to transform the pseudo-distance distributions $C(D)$ into the equivalent hydrogen column density of the dust $\Delta N_{\rm{H}}(D)$, we divided the two sectors into annular regions, with a width of 30$^{\prime\prime}$, ranging from 2$^\prime$ to 12$^\prime$
and extracted the corresponding spectra.
For each spectrum, a specific response matrix was generated.
Using XSPEC \citep{Xspec}, we calculated the expected number of detected counts in each spectrum $C(D)$ for a fixed amount of dust, 
calculating the distance $D$ from Eq.~\ref{eq_distance}, 
where $t$ is the time elapsed between the GRB and the middle of the observation and $\theta$ is the average radius of the spectral extraction region.
For the scattered emission, we adopted the model reported in \citetalias{Tiengo2023} and briefly described in Appendix~\ref{appA}: $\textsc{constant}\,\times\,\textsc{zTBabs}\,\times\,\textsc{TBabs}\,\times\,\textsc{ringscat}\,\times\,\textsc{pegpwrlw}$.
In this case, the \textsc{constant} is the inverse of the exposure time, adjusted by the ratio of the sector's angular dimension to the full circle (105°/360°) and the amount of dust in \textsc{ringscat} is fixed to $\Delta N_{\rm{H}} = 10^{21}\rm{cm^{-2}}$.
All the other spectral parameters, including the GRB fluence of $F_{\rm{GRB}}=2.6\times10^{-3}\,\rm{erg\,cm^{-2}}$, are fixed to the best-fit values reported in \citetalias{Tiengo2023}.
The conversion factor $f(\theta, t)$ for each dataset, 
in units of $10^{21}\rm{cm^{-2}}$, 
is therefore the inverse of the number of expected counts in the 0.7--4 keV energy band, from which the pseudo-distance distributions were extracted.

\end{appendix}

\end{document}